\documentclass[iop]{emulateapj}
\usepackage{apjfonts}
\usepackage{natbib}
\usepackage{mathptmx}
\usepackage{graphicx}
\usepackage{subfigure}

\newcommand{\todo}{\ifmmode {\Huge \bullet} \else {\Huge$\bullet$}\fi}
\newcommand{\Ntot}{40~}
\newcommand{\Nvlt}{29~}
\newcommand{\Nniri}{11~}
\newcommand{\Nphoto}{11~}

\newcommand{\Nukidss}{10~}

\newcommand{\ltsim}{\raisebox{-.5ex}{$\;\stackrel{<}{\sim}\;$}}
\newcommand{\gtsim}{\raisebox{-.5ex}{$\;\stackrel{>}{\sim}\;$}}
\newcommand{\vFWHM}{\ifmmode V_{\mbox{\tiny FWHM}} \else $V_{\mbox{\tiny FWHM}}$ \fi}
\newcommand{\et}{et al.\ }

\newcommand{\kms}{\ifmmode {\rm km\,s}^{-1} \else km\,s$^{-1}$ \fi}

\newcommand{\ergs}{\ifmmode {\rm ergs\,s}^{-1} \else ergs\,s$^{-1}$ \fi}
\newcommand{\ergcms}{\ifmmode {\rm ergs\,cm}^{-2}\,{\rm s}^{-1} \else ergs\,cm$^{-2}$\,s$^{-1}$\fi}
\newcommand{\ergcmsA}{\ifmmode{\rm ergs}\, {\rm cm}^{-2}\,{\rm s}^{-1}\,{\rm\AA}^{-1} \else ergs\, cm$^{-2}$\, s$^{-1}$\, \AA$^{-1}$\fi}
\newcommand{\ergcmsHz}{\ifmmode{\rm ergs\,cm}^{-2}\,{\rm s}^{-1}\,{\rm Hz}^{-1} \else ergs\,cm$^{-2}$\,s$^{-1}$\,Hz$^{-1}$\fi}
\newcommand{\phcms}{\ifmmode {\rm ph\,cm}^{-2}\,{\rm s}^{-1} \else ,ph\,cm$^{-2}$\,s$^{-1}$\fi}
\newcommand{\phcmsA}{\ifmmode {\rm ph\,cm}^{-2}\,{\rm s}^{-1}\,{\rm\AA}^{-1} \else ph\,cm$^{-2}$\,s$^{-1}$\,\AA$^{-1}$\fi}

\newcommand\Msun{\ifmmode M_{\odot} \else $M_{\odot}$\fi}
\newcommand\msun{\ifmmode M_{\odot} \else $M_{\odot}$\fi}
\newcommand\Lsun{\ifmmode L_{\odot} \else $L_{\odot}$\fi}

\newcommand\mpyr{\ifmmode \Msun\,{\rm yr}^{-1} \else $\Msun\,{\rm yr}^{-1}$ \fi}

\newcommand{\Luv}{\ifmmode L_{1450} \else $L_{1450}$\fi}
\newcommand{\Lop}{\ifmmode L_{5100} \else $L_{5100}$\fi}
\newcommand{\Lthree}{\ifmmode L_{3000} \else $L_{3000}$\fi}
\newcommand{\lledd}{\ifmmode L/L_{\rm Edd} \else $L/L_{\rm Edd}$\fi}
\newcommand{\lamLlam}{\ifmmode \lambda L_{\lambda} \else $\lambda L_{\lambda}$\fi}
\newcommand{\lbol} {\ifmmode L_{\rm bol} \else $L_{\rm bol}$\fi}
\newcommand{\llbol}{\ifmmode \log\left(\lbol/\ergs\right) \else $\log\left(\lbol/\ergs\right)$\fi}

\newcommand{\fuv}{\ifmmode f_{\lambda}\left(1450\AA\right) \else $f_{\lambda}\left(1450 {\rm \AA}\right)$\fi}
\newcommand{\fthree}{\ifmmode f_{\lambda}\left(3000\AA\right) \else $f_{\lambda}\left(3000{\rm \AA}\right)$\fi}
\newcommand{\fH}{\ifmmode f_{\lambda}\left(1.65\micron\right) \else
$f_{\lambda}\left(1.65\micron\right)$\fi}

\newcommand{\mbh}{\ifmmode M_{\rm BH} \else $M_{\rm BH}$\fi}
\newcommand{\lmbh}{\ifmmode \log\left(\mbh/\Msun\right) \else $\log\left(\mbh/\Msun\right)$\fi} 
 
\newcommand{\mseed}{\ifmmode M_{\rm seed} \else $M_{\rm seed}$\fi}
\newcommand{\mbul}{\ifmmode M_{\rm Bulge} \else $M_{\rm Bulge}$\fi} 
\newcommand{\mstar}{\ifmmode M_{*} \else $M_{*}$\fi} 
\newcommand{\mhost}{\ifmmode M_{\rm Host} \else $M_{\rm Host}$\fi}
\newcommand{\mm}{\ifmmode M_{*}/M_{\rm BH} \else $M_{*}/M_{\rm BH}$\fi}
\newcommand{\mmwp}{\ifmmode \left(M_{*}/M_{\rm BH}\right) \else $\left(M_{*}/M_{\rm BH}\right)$\fi}
\newcommand{\ml}{\ifmmode M_{*}/L_{*} \else $M_{*}/L_{*}$\fi}
\newcommand{\mlwp}{\ifmmode \left(M_{*}/L\right) \else $\left(M_{*}/L\right)$\fi}
\newcommand{\mlk}{\ifmmode \left(M_{*}/L_{K}\right) \else $\left(M_{*}/L_{K}\right)$\fi}
\newcommand{\sigs}{\ifmmode \sigma_{*} \else $\sigma_{*}$\fi}
\newcommand{\fbol} {\ifmmode f_{\rm bol} \else $f_{\rm bol}$\fi}
\newcommand{\fbolwv} {\ifmmode f_{\rm bol}\left(\lambda\right) \else $f_{\rm bol}\left(\lambda\right)$\fi}
\newcommand{\fbolopt} {\ifmmode f_{\rm bol}\left(5100\AA\right) \else $f_{\rm bol}\left(5100\AA\right)$\fi}
\newcommand{\fboluv} {\ifmmode f_{\rm bol}\left(3000\AA\right) \else $f_{\rm bol}\left(3000\AA\right)$\fi}

\newcommand{\zfpe}{\ifmmode z\simeq4.8 \else $z\simeq4.8$\fi}
\newcommand{\znetprev}{$z\simeq2.4$ and $\simeq3.3$}
\newcommand{\ztpt}{$z \simeq 3.3$}
\newcommand{\ztpf}{$z \simeq 2.4$}
\newcommand{\zsix}{$z \simeq 6.2$}

\newcommand{\hband}{\textit{H}-band}
\newcommand{\kband}{\textit{K}-band}

\newcommand \Hbeta {\ifmmode {\rm H}\beta \else H$\beta$\fi}
\newcommand \hb    {\ifmmode {\rm H}\beta \else H$\beta$\fi}
\newcommand  \mgii  {\ifmmode {\rm Mg}{\textsc{ii}} \else Mg\,{\sc ii}\fi} 
\newcommand  \MgII  {\ifmmode {\rm Mg}\,{\sc ii}\,\lambda2798 \else Mg\,{\sc ii}\,$\lambda2798$\fi}
\newcommand  \civ  {\ifmmode {\rm C}\,{\sc iv} \else C\,{\sc iv}\fi}
\newcommand  \CIV  {\ifmmode {\rm C}\,{\sc iv}\,\lambda1549 \else C\,{\sc iv}\,$\lambda1549$\fi}
\newcommand  \feii  	{Fe\,{\sc ii}}
\newcommand  \feiii 	{Fe\,{\sc iii}}
\newcommand  \oi	{\ifmmode \left[{\rm O}\,{\textsc i}\right] \else [O\,{\sc i}]\fi}
\newcommand  \OI	{\ifmmode \left[{\rm O}\,{\textsc i}\right]\,\lambda6300 \else [O\,{\sc i}]$\,\lambda6300$ \fi}

\newcommand  \oii	{\ifmmode \left[{\rm O}\,{\textsc ii}\right] \else [O\,{\sc ii}]\fi}
\newcommand  \OII	{\ifmmode \left[{\rm O}\,{\textsc ii}\right]\,\lambda3727 \else [O\,{\sc ii}]\,$\lambda3727$ \fi}
\newcommand  \oiii	{\ifmmode \left[{\rm O}\,{\textsc iii}\right] \else [O\,{\sc iii}]\fi}
\newcommand  \OIII	{\ifmmode \left[{\rm O}\,{\textsc iii}\right]\,\lambda5007 \else [O\,{\sc iii}]\,$\lambda5007$\fi}

\newcommand{\lmg}{\ifmmode L\left(\mgii\right) \else $L\left(\mgii\right)$\fi}
\newcommand{\fwmg}{\ifmmode {\rm FWHM}\left(\mgii\right) \else FWHM(\mgii)\fi}
\newcommand{\fwciv}{\ifmmode {\rm FWHM}\left(\civ\right) \else FWHM(\civ)\fi}
\newcommand{\fwhm}{\ifmmode {\rm FWHM} \else FWHM\fi}

\newcommand\nar{{New A Rev.}}%

\shorttitle{BLACK-HOLE MASS AND GROWTH RATE AT $z\simeq4.8$}
\shortauthors{TRAKHTENBROT ET AL.}

\begin{document}

\title{Black-Hole Mass and Growth Rate at $z\simeq4.8$: A Short Episode of Fast Growth Followed by Short Duty Cycle Activity
\footnote{Based on observations collected at the European Organisation for Astronomical Research in the Southern Hemisphere, Chile, as part of programs 081.B-0549, 082.B-0520 and 085.B-0863 
and at the Gemini Observatory, as part of programs GN-2007B-Q-56 and GN-2008B-Q-75.}
}

\author{
Benny Trakhtenbrot\altaffilmark{1},
Hagai Netzer\altaffilmark{1},
Paulina Lira\altaffilmark{2}
and Ohad Shemmer\altaffilmark{3}
}
\altaffiltext{1} {School of Physics and Astronomy and the Wise
  Observatory, The Raymond and Beverly Sackler Faculty of Exact
  Sciences, Tel-Aviv University, Tel-Aviv 69978, Israel. trakht@wise.tau.ac.il}
\altaffiltext{2} {Departamento de Astronom\'ia, Universidad de Chile,
  Camino del Observatorio 1515, Santiago, Chile}
\altaffiltext{3} {Department of Physics, 
University of North Texas, Denton, TX 76203}

\begin{abstract}
We present new Gemini-North/NIRI and VLT/SINFONI \hband\ spectroscopy for a flux limited sample of \Ntot \zfpe\ active galactic nuclei, selected from the Sloan Digital Sky Survey. 
The sample probably contains the most massive active black holes (BHs) at this redshift and spans a broad range in bolometric luminosity, $2.7\times10^{46}< \lbol < 2.4\times10^{47} \ergs$. 
The high-quality observations and the accurate fitting of the \MgII\ line, enable us to study, systematically, the distribution of BH mass (\mbh) and normalized accretion rate (\lledd) at \zfpe.
We find that $10^{8} \ltsim \mbh \ltsim 6.6\times10^{9}\,\Msun$ with a median of $\sim8.4\times10^{8}\,\Msun$.
We also find that $0.2 \ltsim \lledd \ltsim 3.9$ with a median of $\sim0.6$. 
Most of these sources had enough time to grow to their observed mass at \zfpe\ from $z=20$, assuming a range of seed BH masses,  with $\sim40\%$ that are small enough to be stellar remnants. 
Compared to previously studied samples at \znetprev, the masses of the \zfpe\ BHs are typically lower by $\sim0.5$ dex. and their \lledd\ is higher by a similar factor.
The new \zfpe\ sample can be considered as the progenitor population of the most massive BHs at \znetprev. 
Such an evolutionary interpretation requires that the growth of the BHs from \zfpe\ to \ztpt\ and \ztpf\ proceeds with short duty cycles, of about 10-20\%, depending on the particular growth scenario.
\end{abstract}


\section{Introduction}
\label{sec:intro}

The local Universe provides ample evidence for the existence of Super-Massive Black Holes (SMBHs) in the centers of most galaxies. Typical masses are in the range $\mbh\sim10^6-10^9\, \Msun$, with few exceptionally massive objects reaching $\sim10^{10}\, \Msun$.
As first argued by Soltan (1982), the total local BH mass is consistent with the total radiation emitted by accreting SMBHs in active galactic nuclei (AGNs).
%
%
The accumulation of mass onto SMBHs can be traced back through cosmic history by analyzing the redshift-dependent quasar luminosity function.
Such studies suggest that the peak epoch of SMBH growth was at $z\sim2-3$
(e.g., Miyaji \et 2001; Hasinger \et 2005; Silverman \et 2008; Croom \et 2009).
However, this statistical approach does not provide sufficient information about the mass of individual SMBHs at various redshifts. 
A more detailed evolutionary study requires such measurements, in combination with a reliable estimate of the bolometric luminosity (\lbol)
and hence the light-to-mass ratio or, equivalently, the normalized accretion rate $\lledd\equiv\lbol/L_{\rm Edd}$.
%

%
%
Several recent studies suggest that the more massive BHs experience most of their growth at very early epochs ($z\gtsim3$), while those observed as AGNs in the the local Universe tend to have lower \mbh\ (``downsizing''; e.g., Marconi \et 2004; Shankar \et 2009).
A similar effect is suggested for the distributions of \lledd, such that at $z\gtsim2$ most AGNs accrete close to their Eddington limit (Merloni 2004; Shankar \et 2009).
It is also predicted that there should be an anticorrelation between \mbh\ and \lledd\ at all redshifts.
These trends are in general agreement with observations (e.g., McLure \& Dunlop 2004; Netzer \& Trakhtenbrot 2007; Shen \et 2008).
Detailed simulations of galaxy mergers 
predict that the instantaneous AGN luminosity and SMBH accretion rate vary greatly on very short timescales (Di Matteo \et 2005; Hopkins \et 2006; Sijacki \et 2007).
The overall BH accretion period following major mergers lasts $\sim1\, {\rm Gyr}$, out of which the central source would appear as a luminous, unobscured AGN for at most a $few \times\,100\, {\rm Myr}$.
\mbh\ may grow by as much as a factor $\sim1,000$ during such mergers. 
The merger history of SMBH hosts can also be traced in cosmological simulations of structure formation, by following halo merger trees (e.g., Volonteri et 2003). 
Almost all these models require (or assume) that at $z\gtsim3$ most AGN would accrete close to, or indeed at their Eddington limit. 
The fast growth has to last almost continuously from very early epochs ($z\sim20$) and involve massive seed BHs ($\mseed\gtsim10^3\,\Msun$), to account for the very massive BHs detected at $z\sim6$ (e.g., Fan \et 2006; Volonteri 2010, and references therein).
The combination of structure formation models predictions at $z\sim3-6$ with the recently observed high clustering of high-redshift luminous AGNs (Shen \et 2007) suggests that the typical duty cycles $-$ the fraction of the total time involving fast accretion $-$ should remain above 0.5 and probably reach unity (e.g. White \et 2008; Wyithe \& Loeb 2009; Shen \et 2010; Shankar \et 2010a; Bonoli \et 2010).  
These studies use various prescriptions to link AGNs with their dark matter halos. 
At lower redshifts (i.e. $z\ltsim2$), the accretion can be more episodic with a  typical duty cycle between $\sim10^{-3}$ and $\sim0.1$ (e.g., Marconi \et 2004; Merloni 2004; Shankar \et 2009)  
A comprehensive, up-to-date review of many of these issues is given in Shankar (2009).
In order to test these scenarios, \mbh\ and \lledd\ ought to be measured in large, representative samples.
For unobscured, type-I AGNs, this is usually achieved by using ``single epoch'' (or ``virial'') \mbh\ determination methods, which are based on the results of long-term reverberation mapping campaigns.
These methods are based on estimating the size of the broad line region (BLR) 
and involve an empirical relation of the form $R_{\rm BLR} \propto \left(\lambda L_{\lambda}\right)^{\alpha}$, where $\lambda L_{\lambda}$ is the monochromatic luminosity in a certain waveband. 
Combining $R_{\rm BLR}$ with the assumption of virialized motion of the BLR gas, 
we get $\mbh=f G^{-1} L^{\alpha} V_{\rm BLR}^2$, where $f$ is a geometrical factor of order unity
(e.g., Kaspi \et 2000; Vestergaard \& Peterson 2006; Bentz \et 2009).
The effect of radiation pressure force on 
such estimates is still a matter of some discussion (Marconi \et 2008),  
but recent work suggests it is not very important (Netzer 2009a; Netzer \& Marziani 2010).
Single epoch mass estimate methods based on the \hb\ and \MgII\ lines (e.g., Kaspi \et 2005 and McLure \& Dunlop 2004, respectively) were used to estimate \mbh\ up to $z\simeq2$ in large optical surveys (e.g., Corbett \et 2003; McLure \& Dunlop 2004; Netzer \& Trakhtenbrot 2007; Fine \et 2008; Shen \et 2008).
Much smaller samples of $z>2$ sources were studied by observing \hb\ or \mgii\ in one of the NIR bands (Shemmer \et 2004, hereafter S04; Kurk \et 2007, hereafter K07; Netzer \et 2007, hereafter N07; Marziani \et 2009; Willott \et 2010, hereafter W10).
\mbh\ can also be estimated from the broad \CIV\ line, using specifically calibrated relations (e.g., Vestergaard \& Peterson 2006). This would potentially enable the study of large samples of AGN at high redshifts.
However, there is clear evidence that \civ-based estimates of \mbh\ are unreliable. 
Baskin \& Laor (2005) found that the \civ\ line is often blue-shifted with respect to the AGN rest-frame, which suggests that the dynamics of the \civ-emitting gas may be dominated by non-virial motion.
Several studies of large samples clearly demonstrate that the relation between the widths of the \civ\ line and of the lower ionization lines (\hb\ and \mgii) is weak and shows considerable scatter (e.g., Shen \et 2008; Fine \et 2010) that is inconsistent with the virial assumption used in such mass estimators.
Finally, our own study (N07) of luminous \znetprev\ AGNs shows a complete lack of correlation between \civ-based and \hb-based estimates of \mbh.
These discrepancies become crucial at high redshifts and large \mbh\ and lead to the conclusion that only \hb-based and \mgii-based
mass estimates are reliable enough to infer the properties of such sources.
Our previous project (S04 \& N07) presented the largest sample of \znetprev\ type-I AGNs for which \mbh\ and \lledd\ were reliably measured using NIR \hb\ spectroscopy.
The distribution of \lledd\ at \znetprev\ was found to be broad, and about half of the sources had $\lledd<0.2$, inconsistent with several of the models mentioned above.
In particular, the typically low accretion rates and the very high masses (up to $\lmbh\simeq10.5$) also mean that $\sim60\%$ of the sources did not have enough time to grow to the observed \mbh\ by continuous accretion at the observed rates.
These findings suggest that an epoch of faster SMBH growth must have occurred, for most objects, at $z>3.5$.
In order to probe such redshifts, 
the \mgii\ line must be observed in either the $H$ or the $K$-bands. 
Practically, this corresponds to focusing on  \zfpe\ or \zsix\ sources.

In this paper we present a systematic study of \mbh\ and \lledd\ in a large, well-defined sample of \zfpe\ type-I AGNs. 
This is based on new \hband\ spectroscopic observations, which enable the measurement of the \mgii\ line.
We describe the sample selection and the observations in \S\ref{sec:sample_obs} and the way we deduced \mbh\ and \lledd\ in \S\ref{sec:fit_mbh_lledd}. 
The main results of these measurements are presented in \S\ref{sec:results} and discussed in \S\ref{sec:discussion} where we compare these results to those of other high-redshift samples.
The main findings are summarized in \S\ref{sec:summary}.
Throughout this work we assume a standard $\Lambda${\sc CDM} cosmology with 
$\Omega_{\Lambda}=0.7$, $\Omega_{M}=0.3$ and $H_{0}=70$\,\kms\,Mpc$^{-1}$.

\section{Sample Selection, Observations and Data Reduction}
\label{sec:sample_obs}

\begin{deluxetable*}{lcclccc}

\tablecolumns{5}
\tablewidth{0pt}
\tablecaption{Observation Log \label{tab:obs_log}}
\tablehead{
\colhead{Object ID (SDSS~J)} &
\colhead{$z_{\rm SDSS}$\tablenotemark{a}} &
\colhead{Instrument} &
\colhead{Obs. Date} &
\colhead{Total Exp.} &
\multicolumn{2}{c}{$H$-band magnitude} \\
  &  &  &  & Time (sec) & spectro.\tablenotemark{b} & imaging\tablenotemark{c}
}
\startdata
000749.17$+$004119.4 & 4.837 & SINFONI  & 6 \& 29 Jun. 2008  		& 4200 & 18.59 &  (18.1)	\\
003525.28$+$004002.8 & 4.757 & SINFONI  & 29 \& 30 Jun. 2008  		& 4200 & 17.82 &  (18.0)	\\
021043.15$-$001818.2 & 4.733 & NIRI  	& 4 Sep. 2007   		& 6380 & 18.73 &  (17.8)	\\
033119.67$-$074143.1 & 4.738 & SINFONI 	& 1 Oct. 2008   		& 2400 & 17.49 &  	\\
075907.58$+$180054.7 & 4.861 & NIRI 	& 18 Oct. \& 11 Nov. 2007  	& 8100 & 17.31 &  	\\
080023.03$+$305100.0 & 4.687 & NIRI 	& 7 Dec. 2008   		& 6160 & 17.06 &  	\\
080715.12$+$132804.8 & 4.874 & NIRI 	& 21 Dec. 2008   		& 4590 & 17.52 &  	\\
083920.53$+$352457.6 & 4.777 & NIRI 	& 15 Dec. 2008   		& 7290 & 18.26 &  	\\
085707.94$+$321032.0 & 4.776 & NIRI 	& 6 Jan. 2008   		& 6490 & 17.06 &  	\\
092303.53$+$024739.5 & 4.660 & SINFONI 	& 23 Dec. 2008 \& 1 Jan. 2009 	& 6000 & 18.44 &  (18.4)	\\
093508.50$+$080114.5 & 4.699 & SINFONI 	& 20 Dec. 2008   		& 2400 & 17.89 & 17.9 	\\
093523.32$+$411518.7 & 4.836 & NIRI 	& 7 \& 12 Dec. 2008  		& 6440 & 17.36 &  	\\
094409.52$+$100656.7 & 4.748 & SINFONI 	& 23 Mar. 2010   		& 2400 & 17.77 &  	\\
101759.64$+$032740.0 & 4.917 & SINFONI 	& 28 Nov. 2008 \& 2 Jan. 2009 	& 6000 & 18.63 &  (18.9)	\\
105919.22$+$023428.8 & 4.735 & SINFONI 	& 17 Feb. 2009   		& 2400 & 17.75 & 18.0 	\\
111358.32$+$025333.6 & 4.882 & SINFONI 	& 6 Apr. 2008   		& 3000 & 17.9 &  	\\
114448.54$+$055709.8 & 4.793 & SINFONI 	& 6, 8 \& 22 Apr. 2008  	& 8100 & 18.56 & 18.6 	\\
115158.25$+$030341.7 & 4.698 & SINFONI 	& 22 Apr. 2008   		& 6000 & 18.94 & 	\\
120256.44$+$072038.9 & 4.785 & SINFONI 	& 3 \& 6 Jan. 2009  		& 4800 & 18.15 & 18.0 	\\
123503.04$-$000331.6 & 4.723 & SINFONI 	& 4 May. 2008   		& 8700 & 18.46 & 	\\
130619.38$+$023658.9 & 4.852 & SINFONI 	& 6 Feb. 2009   		& 3300 & 16.81 & 16.9 	\\
131737.28$+$110533.1 & 4.810 & SINFONI 	& 20 \& 21 Jan. 2009  		& 4500 & 18.01 &  (17.9)	\\
132110.82$+$003821.7 & 4.716 & SINFONI 	& 7 \& 9 Apr. 2008  		& 3000 & 18.4 &  	\\
132853.67$-$022441.7 & 4.695 & SINFONI 	& 8 Feb. 2009   		& 2400 & 18.07 &  	\\
133125.57$+$025535.6 & 4.737 & SINFONI 	& 14 \& 26 Apr. 2008  		& 4800 & 18.67 & 18.7 	\\
134134.20$+$014157.8 & 4.670 & SINFONI 	& 8 Feb. 2009   		& 1800 & 17.02 &  	\\
134546.97$-$015940.3 & 4.714 & SINFONI 	& 26 Apr. 2008   		& 1800 & 18.38 &  	\\
140404.64$+$031404.0 & 4.870 & SINFONI 	& 28 Mar. 2010   		& 4500 & 17.71 & 17.7 	\\
143352.21$+$022714.1 & 4.721 & SINFONI 	& 29 Mar. 2010   		& 600  & 16.7 &  	\\
143629.94$+$063508.0 & 4.850 & SINFONI 	& 28 Mar. 2010   		& 2400 & 17.79 & 18.0 	\\
144352.95$+$060533.1 & 4.879 & SINFONI 	& 4 Apr. 2010   		& 4800 & 18.56 & 18.9 	\\
144734.10$+$102513.2 & 4.686 & SINFONI 	& 7 \& 8 Apr. 2010  		& 5400 & 18.78 &  (18.6)	\\
151155.98$+$040803.0 & 4.686 & SINFONI 	& 4 Apr. 2010   		& 2400 & 17.99 & 17.8 	\\
161622.11$+$050127.7 & 4.872 & SINFONI 	& 8 Apr. 2010   		& 900  & 16.89 & 17.3 	\\
165436.86$+$222733.7 & 4.678 & NIRI 	& 19 Aug. 2007   		& 3480 & 17.66 &  	\\
205724.15$-$003018.0 & 4.663 & NIRI 	& 12 Aug. 2008   		& 5800 & 16.77 &  	\\
220008.66$+$001744.8 & 4.818 & NIRI 	& 12 Oct. 2007   		& 6160 & 17.51 &  (17.5)	\\
221705.72$-$001307.7 & 4.689 & SINFONI 	& 19 Apr. 2008   		& 3300 & 18.12 &  (18.4)	\\
222509.16$-$001406.8 & 4.888 & NIRI 	& 3 Sep. 2007   		& 6380 & 17.15 &  (17.1)	\\
224453.06$+$134631.8 & 4.657 & SINFONI 	& 6 \& 8 Jul. 2010 		& 4800 & 18.67 &  	\\
\enddata
\tablenotetext{a}{Redshift obtained from the SDSS archive, based on
  rest-frame UV emission lines.}
\tablenotetext{b}{$H$-band (Vega) magnitude derived from the calibrated spectra by synthetic photometry.}
\tablenotetext{c}{$H$-band (Vega) magnitude derived from direct CTIO/ISPI imaging, or from the UKIDSS database (in parentheses).}
\end{deluxetable*}

\subsection{Sample selection}
\label{sec_sub:sample}

Sources were selected to allow measurement of the \MgII\ emission line and the continuum flux at 3000\AA\, ($F_{3000}$), thus providing reliable \mbh\ and \lledd\ estimates.
The sources were selected from the sixth data release (DR6; Adelman$-$McCarthy \et 2008) of the Sloan Digital Sky Survey (SDSS; York \et 2000).
We first limited our search to ``QSO''-class objects with $z\sim4.65-4.95$, so that the \mgii\ line and $F_{3000}$ could be observed within the \textit{H}-band. 
The completeness rate of SDSS spectroscopy for QSOs at this redshift is above $98\%$, down to a limiting magnitude of $i\simeq20$ (Richards \et 2006a). 
This initial search resulted in 177 objects.
Next, we verified that no broad absorption features are present near the SDSS-observed \civ\ lines, since these are also expected to appear in the \mgii\ profile.
This reduced the list of candidates to 129 sources.
To estimate the \hband\ flux of the targets, we extrapolated the rest-frame flux density at 1450\AA\ to the \textit{observed}-frame flux density at $1.65\mu{\rm m}$, using the $f_\nu \propto \nu^{-0.44}$ spectral energy distribution (SED) presented by Vanden Berk \et (2001).
%
We applied a flux limit of
$\fuv \gtsim 6 \times 10^{-18}\, \ergcmsA$, 
which translates to  
$\fH  \gtsim 2.1\times 10^{-18}\, \ergcmsA$,   
to include only sources which would provide NIR spectra with $S/N\gtsim5-10$ within reasonable exposure times. 
This formal flux limit translates to $\lbol \simeq 5\times10^{46}\, \ergs$ (see below).
Our flux limit results in the omission of only 6 candidates, leaving 123 sources.
Different observational constraints (a combination of declination, observational seasons etc.; see \S\S\ref{sec_sub:spec_obs}), forced us to observe only \Ntot of these 123 candidates.
We verified that the distribution of \fuv\ for the observed sample is very similar to that of the flux-limited sample of 123 sources, as well as the entire initial sample of 177 SDSS objects.
In particular, we find 4 sources ($\sim11\%$ of the observed sample) with $\fuv\ltsim 7.5\times 10^{-18}\, \ergcmsA$ and 15 sources ($\sim43\%$) with $\fuv\ltsim 1.5\times 10^{-17} \ergcmsA$. 
Given this and the high level of completeness in the SDSS, we consider this sample to be representative of the complete sample of \zfpe\ luminous AGNs.

Several studies suggest that the mass and accretion rate evolution of SMBHs may be connected with their radio properties (see, e.g., McLure \& Jarvis 2004; Shankar \et 2010b, but also Woo \& Urry 2002).
To determine the radio properties of the \zfpe\ sources, we utilized the cross-matched catalog Faint Images of the Radio Sky at Twenty-Centimeters radio survey (FIRST; Becker \et 1995; White \et 1997). 
Due to the high redshift of our sample, and the relatively low sensitivity of FIRST, most candidates have only upper limits on their radio fluxes, and thus upper limits on the radio-loudness ($R_{\rm L}\equiv \frac{ f_{\nu}\left(5\, {\rm GHz}\right)}{f_{\nu}\left(4400\AA\right)}$; Kellermann \et 1989).
For those sources, we estimated $R_{\rm L}$ based on a $3\sigma$ FIRST upper flux limit of $f_{\nu}\left(5 {\rm GHz}\right)=0.6\, {\rm mJy}$ and assuming a radio SED of $f_\nu \propto \nu^{-0.8}$. 
$f_{\nu}\left(4400\AA\right)$ was estimated from \fuv\ and the Vanden Berk et (2001) template. 
All but five observed targets (see Table~\ref{tab:obs_log}) have upper limits on  $R_{\rm L}$ which are below $\sim40$.
Two of the remaining sources (J0210-0018 \& J1235-0003) have firm FIRST detections ($f_{\nu}\left[5 {\rm GHz}\right]=9.75$ and  $18.35\, {\rm mJy}$, respectively) and are thus considered as radio loud AGNs ($R_{\rm L}\simeq 104$ and $1080$, respectively). 
Three additional targets were not observed by FIRST and thus have no viable radio data.
In what follows, we verified that the inclusion or exclusion of these 5 sources does not significantly affect our results.
In summary, the \Ntot \zfpe\ AGNs presented here comprise a flux limited sample 
which represents a large fraction of all such SDSS sources.
In terms of bolometric luminosity, it is complete down to $\lbol \simeq5\times10^{46}\ \ergs$ and most of the sources are not radio-loud.
The basic properties of the \Ntot sources are given in Table~\ref{tab:obs_log}. 

\subsection{Spectroscopic observations and reduction}
\label{sec_sub:spec_obs}

\begin{figure*}
\includegraphics[width=18cm]{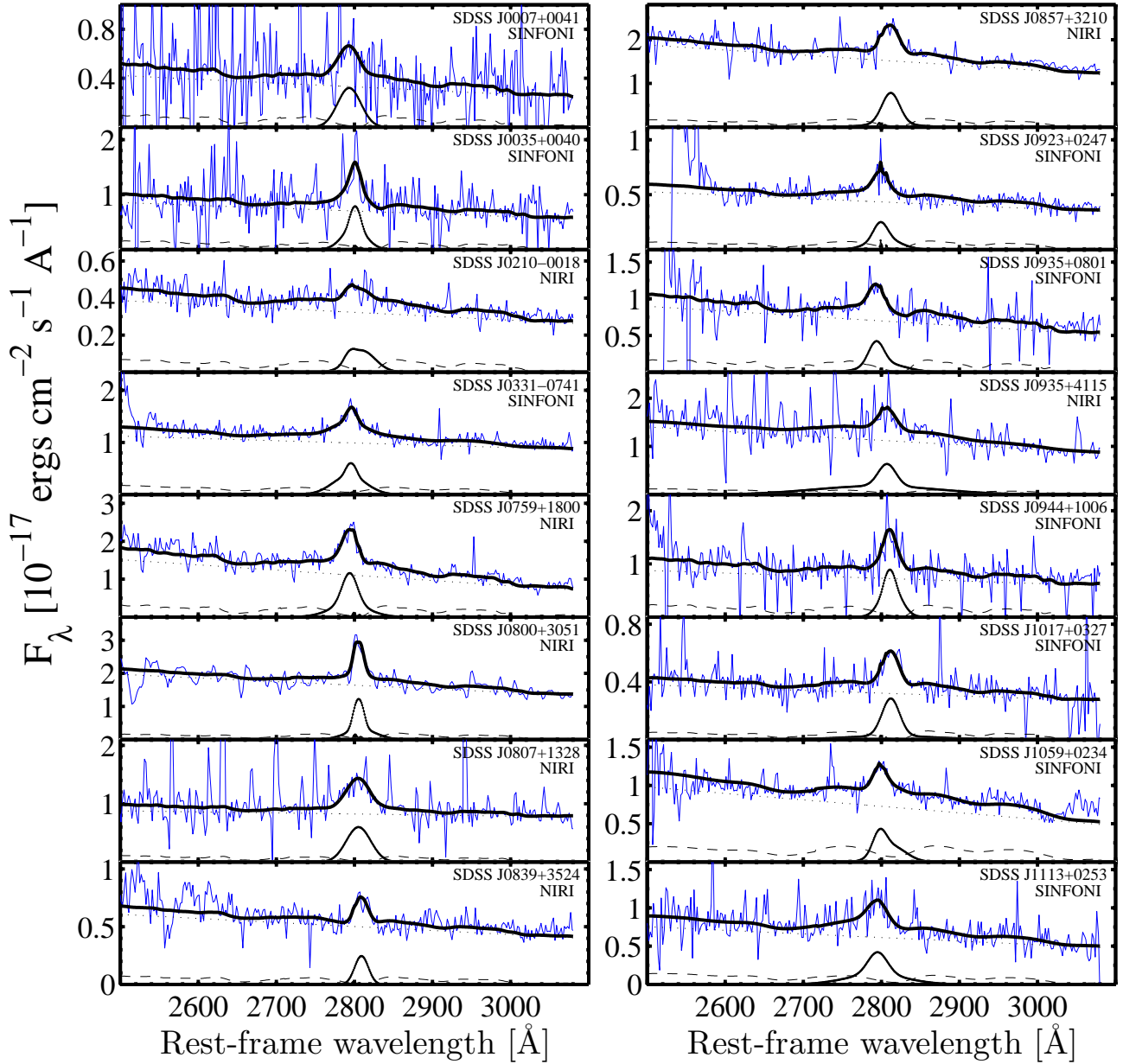}
\caption{Rest-frame spectra of the \zfpe\ sources under study.
The spectra are ordered by increasing right ascension. 
In each panel, we show the observed flux density (thin blue line) 
and the best-fit model (thick black line), which is composed of a continuum component (dotted line), an \feii\ \& \feiii\ emission complex (thin dashed curve), and the total (two-component) \MgII\ line (thin black lines).}
\label{fig:spectra}
\end{figure*}

\addtocounter{figure}{-1}
\begin{figure*}
\includegraphics[width=18cm]{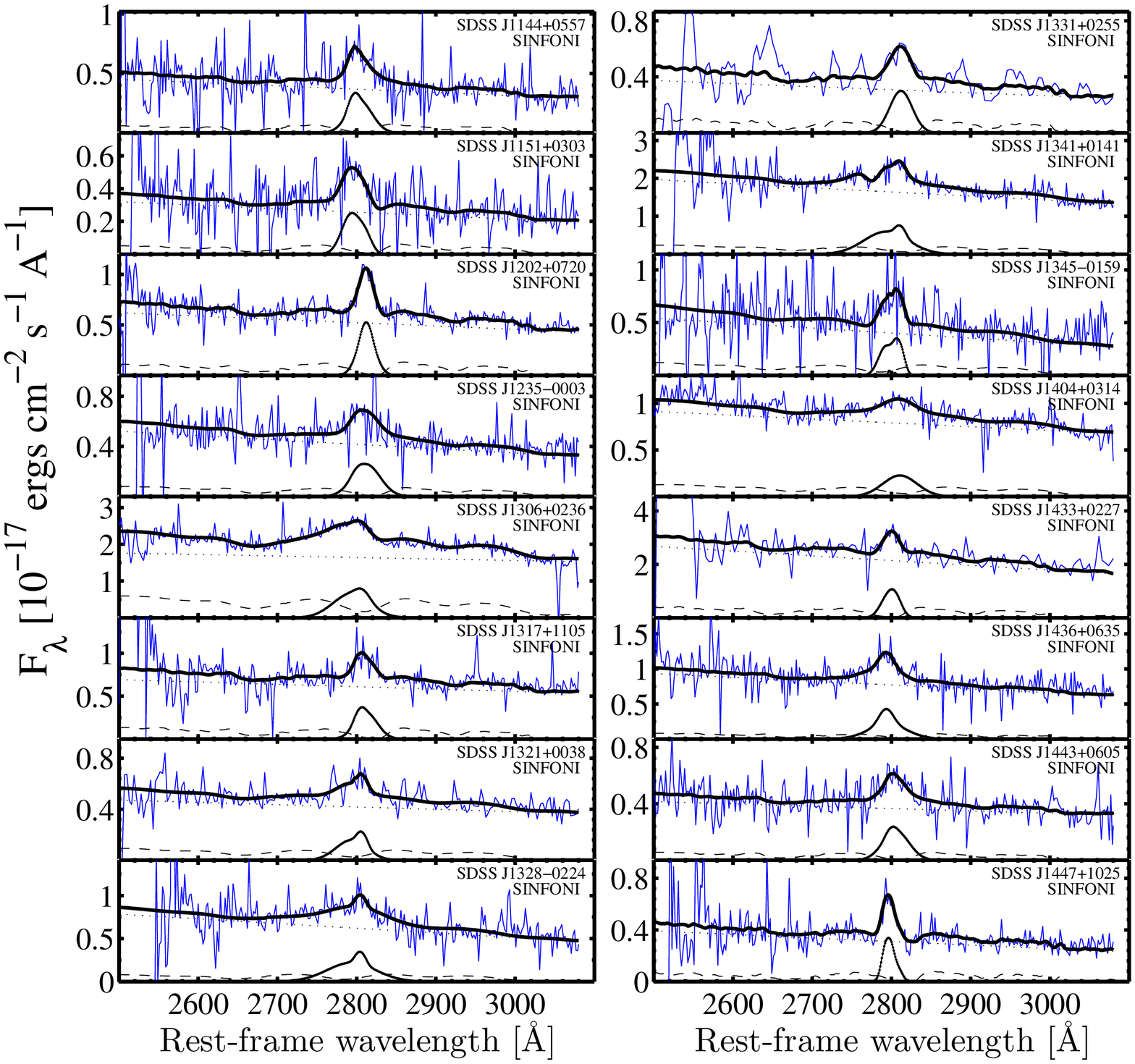}
\caption{-continued.}
\end{figure*}

\addtocounter{figure}{-1}
\begin{figure*}
\includegraphics[width=18cm,clip=true,trim=0 0 0 10cm]{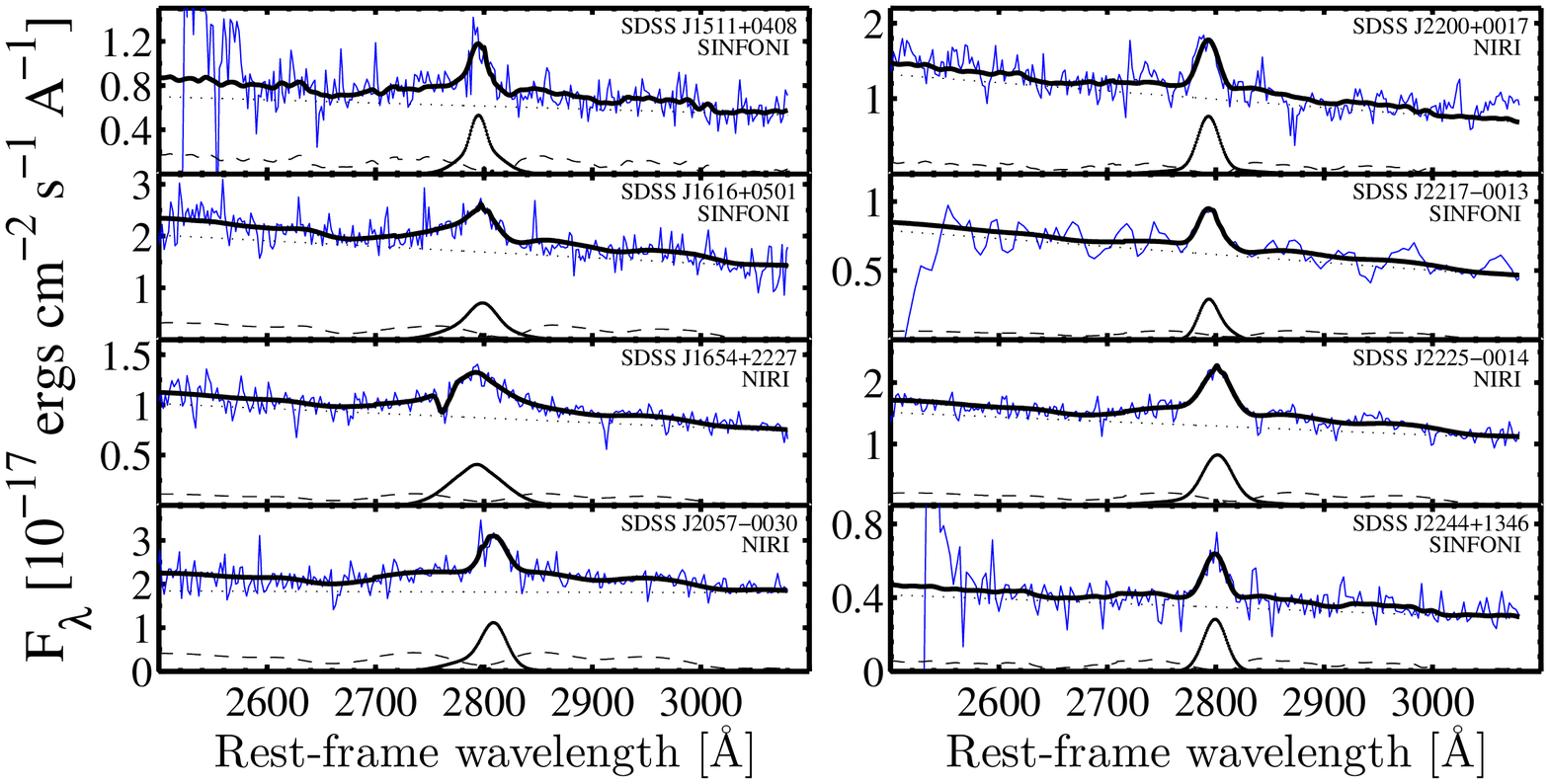}
\caption{-continued.}
\end{figure*}

The \hband\ spectra of our \zfpe\ sample were obtained using the Gemini-North and the Very Large Telescope (VLT) observatories.
The campaign was split over several semesters, with brighter targets predominantly observed with the Near Infra-Red Imager instrument (NIRI; Hodapp \et 2003) on Gemini-North (as part of programs GN-2007B-Q-56 and GN-2008B-Q-75) and fainter ones with the more sensitive SINFONI instrument (Eisenhauer \et 2003) on the VLT-UT4 (as part of programs 081.B-0549, 082.B-0520 and 085.B-0863).
The log of observations is given in Table~\ref{tab:obs_log}.
All the observations were performed in queue/service modes, requested not to exceed airmasses of $\sim1.7$ and seeing of $\sim1\arcsec$, during clear skies and ``gray-time'' lunar phase.
The \Nniri Gemini-N/NIRI targets were observed through a $0.75\arcsec\times110\arcsec$ slit and the G5203 grism at the f/6 setup, providing $R\sim520$.
The sub-integrations were of $\sim300$ sec., with dithers of 6\arcsec along the slit after each sub-integration.
%
The \Nvlt VLT/SINFONI targets were observed through the $8\arcsec\times8\arcsec$ FOV (a.k.a ``250 mas/spaxel'')
We used SINFONI's ``H+K'' mode, since the broad \MgII\ line can be well-resolved with the resulting $R\sim1500$. 
The \kband\ spectra assisted in determining the continuum flux of the SINFONI targets.
Sub-integrations were of 150 sec. or 300 sec. and (diagonally) dithered by $\sim6\arcsec$ across the FOV.
Despite this type of dithering, the spectroscopically resolved OH sky emission contributes a major source of noise to our reduced data (see below).
In both observatories, telluric standards of spectroscopic types B, A and G, chosen to have similar airmasses, were observed immediately before or after the science targets.

The reduction of the raw data was carried out using the standard pipelines of the respective facilities.
We used the {\sc gemini v.1.8} package 
in IRAF to co-align and combine the sub-integrated frames while subtracting the sky emission.
The one-dimensional spectra were extracted through a typical aperture of 20 pixels which correspond to $\sim2.3\arcsec$.
The {\sc SINFONI v.2.0.5} pipline (under the {\sc Esorex} environment) was used to extract ``3D data-cubes'' from each sub-integrated frame, which include an individual wavelength calibration. 
The data-cubes were then co-aligned and combined, while the sky emission was estimated from the appropriate dithered pointings. 
The combined data-cube of each source was examined to estimate the actual PSF, and an appropriate circular aperture was used to extract the 1-d spectra. 
%
Due to the varying flexture of the instrument during the long (1hr) SINFONI Observing Blocks, there is a noticeable ``wavelength flexture'' effect in the extracted 1-d spectra. 
This effect results in P-Cygni-like spectral features and can be treated separately (see Davies 2007).
We note that the observable properties critical to this work (i.e. \Lthree\ and FWHM[\mgii]) are not very sensitive to such narrow ($\ltsim400\,\kms$) OH-originated features.
The spectra of the standard stars were reduced and extracted in the same manner as above, using the same apertures as those used for the science targets.
The published spectral types of these stars were used to correct for the instrumental response of each science observations and their published (2MASS; Cutri \et 2003) magnitudes were used to flux-calibrate the science spectra.
The typical photometric error associated with the 2MASS magnitudes is $\ltsim10\%$.

\subsection{Photometric observations}
\label{sec_sub:photo_obs}

Wide-field $H$ and $K$-band imaging for \Nphoto of the \zfpe\ sources was obtained at the CTIO Blanco telescope, with the ISPI instrument (van der Bliek \et 2004).
We aimed at having at least 3 bright 2MASS stars observed in the same field each of the science targets.
Each field was imaged to achieve S/N\gtsim20 for the science targets, through a series of dithered short exposures, to avoid saturation.
The CTIO observations were carried out with mostly clear conditions, with seeing of $\sim1\arcsec$, during a single run on the night of March 4th., 2010.
%
%
The raw imaging data was reduced following standard procedures, including bad pixel removal, dark and flat calibration and co-alignment of the sub-exposures.
Aperture photometry was performed over the science targets and the several visible 2MASS stars in each field.
Whenever possible, we used only stars with best-quality 2MASS photometry for the relative scaling of the science target fluxes.
The magnitudes thus obtained are also listed in Table~\ref{tab:obs_log}, for the relevant sources.
Since the spectroscopy and imaging observations were not simultaneous, there exists a real uncertainty regarding possible flux variability of our \zfpe\ sources. 
Our targets are relatively luminous, and thus are not expected to have large variability amplitudes (e.g. Bauer \et 2009 and references therein).
We compared the photometric and the spectroscopic fluxes, achieved by synthetic photometry of the SINFONI \& NIRI spectra. 
The median discrepancy is $\ltsim0.09$ mags, in the sense that the (later observed) photometric fluxes are on average fainter than the spectroscopy-deduced fluxes.
The standard deviation is close to 0.2 mags. 
Photometric fluxes for \Nukidss additional sources were obtained through the fourth data release of the UKIRT Infrared Digital Sky Survey (UKIDSS/DR4; Lawrence \et 2007). 
The photometry of these sources is also consistent with the fluxes deduced from our spectrophotometric calibration. Excluding one significant outlier (J0210-0018), the mean deviation between the two methods is, again, close to 0.2 mags. 
With this evidence of little flux variation, and reliable spectrophotometry for a large fraction ($\sim50\%$) of arbitrary selected sources, 
we conclude that the accuracy of our spectrophotometric flux calibration is high, with uncertainties of about $10-15\%$.
In order to correctly probe the instantaneous emission from the sources, we thus adopt the spectrophotometric fluxes to measure luminosities, even for those sources which were photometrically observed.
The full set of calibrated spectra is presented in Figure~\ref{fig:spectra}.

\section{Line Fitting and \mbh\ \& \lledd\ Determination}
\label{sec:fit_mbh_lledd}

\subsection{Line fitting}
\label{sec_sub:fit}

To determine \mbh\ and \lledd\ we fit the observed spectra with a model which combines a continuum component, \feii and \feiii\ lines, and the two doublet \mgii\ lines. 
The procedure used here is similar to the one presented in several other studies (e.g. Shen \et 2008 and references therein).
It is based on a new code that was developed and extensively tested on a large sample of $0.5\ltsim z \ltsim 2$ SDSS type-I AGNs, and will be described in detail in a future publication (Trakhtenbrot \et 2011).
In particular, this large sample was used to verify that important quantities such as narrow emission lines and bolometric corrections, are consistent with those derived from the well-studied \hb-\OIII\ emission region (rest-frame wavelengths of $4600-5100$\AA). 
This was done using a sub-sample of $\sim5,000$ sources at $0.5\ltsim z \ltsim 0.75$, where the SDSS spectra show both the \hb\ and \mgii\ lines.

The fitting of the \zfpe\ AGNs was preformed individually, verifying a satisfactory match between the data and the model. 
As a first step, a linear pseudo-continuum is fit to the flux around 2655 and 3020\AA.
Next, we fit the \feii\ and \feiii\ emission complexes redward and blueward of \mgii.
For this we use a template made of the composite prepared by Vestergaard \& Wilkes (2001) that was supplemented by several predicted \feii\ lines 
(kindly provided by G.~Ferland). 
These are mainly \feii\ emission lines which coincide in wavelength with the \mgii\ line, where the observationally-based template of Vestergaard \& Wilkes (2001) is incomplete.
This part of the template is very similar to the models presented in Sigut \& Pradhan (2003; their Fig.~13) and in Baldwin \et (2004; their Fig.~5). 
In both cases, the \feii\ flux under the \mgii\ line is dominated by the red extension of the $\sim2750$\AA\ emission complex.
The additional \feii\ flux tends to flatten in sources with very broad lines, but the overall effect on the \mgii\ line fitting is very small. 
We consider this template to be more reliable than the addition of constant flux under the \mgii\ line, adopted by Kurk \et (2007) and Fine \et (2008). 
Several other studies, such as Salviander \et (2007) and Shen \et (2008), use templates very similar to ours, following the models in Sigut \& Pradhan (2003).
Our basic iron template is based on a single line profile with ${\rm FWHM}\simeq1,175\, \kms$. 
To account for the range of observed line widths, we created a grid of broadended Fe templates, by convolving the basic template with single Gaussian profiles of varying width, ranging from ${\rm FWHM}=1,200$ to $10,000\, \kms$.
We note that the templates that correspond to ${\rm FWHM} \gtsim 4,000\, \kms$ lack almost any distinguishable emission features.
The grid of broadened Fe templates is fitted to the continuum-subtracted spectra over the rest-frame wavelength regions of $2600-2700$\AA\ and $2900-3030$\AA, and the best-fit template is chosen by standard $\chi^2$ minimization. 
We allow the Fe template to be shifted with respect to the systemic redshift and for its width to differ from that of the \mgii\ line. 
After the subtraction of the best-fit Fe template, the pseudo-continuum is re-fitted and we iterate the Fe-fitting process once again. This is done to ensure the convergence of the best-fit pseudo-continuum and Fe lines.
The final measure of the pseudo-continuum flux, \fthree, provides the continuum luminosity \Lthree\ which is further used to calculate \mbh\ and \lledd.
It is important to note that the \Lthree\ thus obtained is \textit{not} the intrinsic, underlying AGN continuum since it also includes a contribution from the Balmer continuum emission, that is not accounted for by the fitting procedure.

Once the pseudo-continuum and Fe contributions are subtracted, we fit the \mgii\ line itself.
The model for the line consists of 3 Gaussians, two broad and one narrow component for each of the \mgii\ doublet lines.
The intensity ratio of the doublet components is fixed to 1:1, suitable for optically-thick lines.
The broad components are limited to line widths in the range $1,200<{\rm FWHM}<10,000\, \kms$, 
while the width of the narrow components is bound to $300<{\rm FWHM}<1,200\, \kms$.
Although the relative contribution of the narrow components is very small ($<10\%$), the procedure reproduces well the NLR width. 
This was verified by comparing the NLR \fwhm\ resulting from the (separate) \mgii\ and \hb\ fitting for the test sample of $\sim5,000$ sources at $0.5\ltsim z \ltsim 0.75$ mentioned above. 
We have also included a single absorption feature (also a Gaussian) to account for the frequent appearance of blue-shifted \mgii\ troughs. 
The best-fit model is used to measure the FWHM of the total broad \textit{single} \mgii\ line and the total luminosity of both \mgii\ components.
We stress that in earlier studies, the line width was calculated using the combined doublet profile, 
which has implications to the deduced \mbh\ (see \S\S\ref{sec_sub:dis_compare}). 
The best-fit continuum and \mgii\ line parameters are given in Table~\ref{tab:line_pars}.
The uncertainties that also appear in Table~\ref{tab:line_pars} reflect the true uncertainties associated with the spectroscopic reduction and line-fitting procedure. 
These were estimated from varying several of the parameters involved in these processes
\footnote{For example, varying flux calibrations for sources which were observed during several nights, the shape and normalization of the \feii\ template and the possible identification of absorption features.} 
and examining the range of possible outcomes. 
We consider these uncertainties to be much more realistic than those derived solely from the flux noise.

\subsection{Estimating \mbh\ and \lledd}
\label{sec_sub:mbh_lledd}

The \lbol\ and \mbh\ estimates used in this work are based on the measured \Lthree\ and \fwmg.
For the latter we adopt the McLure \& Dunlop (2004) relationship:
\begin{equation}
  \mbh=3.2\times 10^6 \left[\frac{\Lthree}{10^{44}\,\ergs}\right]^{0.62}
\left[\frac
    {\fwmg}{10^3 \,\kms}\right]^2 \,\,\Msun \, .
\label{eq:Mbh_MD04}
\end{equation}

\begin{deluxetable*}{lcccccccccc}

\tablecolumns{11}
\tablewidth{0pt}
\tablecaption{Observed and Derived Properties \label{tab:line_pars}}
\tablehead{
  \colhead{Object ID (SDSS~J)}  &
  \colhead{$z_{\rm SDSS}$}  &
  \colhead{$z_{\mgii}$\tablenotemark{a}}  &
  \colhead{$\log$ \Luv \tablenotemark{b}} &
  \colhead{$\log$ \Lthree} &
  \colhead{$L$-qual. \tablenotemark{c}}  &
  \colhead{$\log$ \lbol}  &
  \colhead{\fwmg} &
  \colhead{FWHM-qual. \tablenotemark{d}}  &
  \colhead{$\log$ \mbh} &
  \colhead{$\log$ \lledd}  \\
    &  &  & (\ergs) & (\ergs) &  & (\ergs) & (\kms) &  & (\Msun) &  }
\startdata
J000749.17+004119.4 & 4.837 & 4.786 & 46.39 & 46.04 & 3 & 46.57 & 3665 & 3 &  8.90 & -0.51 \\ 
J003525.28+004002.8 & 4.757 & 4.759 & 46.35 & 46.38 & 3 & 46.91 & 1805 & 3 &  8.49 &  0.24 \\ 
J021043.15-001818.2 & 4.733 & 4.713 & 46.66 & 46.04 & 2 & 46.56 & 4583 & 3 &  9.09 & -0.70 \\ 
J033119.67-074143.1 & 4.738 & 4.729 & 46.76 & 46.55 & 1 & 47.09 & 2360 & 1 &  8.83 &  0.08 \\ 
J075907.58+180054.7 & 4.861 & 4.804 & 46.46 & 46.54 & 1 & 47.07 & 2717 & 1 &  8.95 & -0.05 \\ 
J080023.03+305100.0 & 4.687 & 4.677 & 46.82 & 46.73 & 1 & 47.26 & 1404 & 1 &  8.49 &  0.59 \\ 
J080715.12+132804.8 & 4.874 & 4.885 & 46.71 & 46.53 & 2 & 47.07 & 3837 & 3 &  9.24 & -0.35 \\ 
J083920.53+352457.6 & 4.777 & 4.795 & 46.70 & 46.24 & 2 & 46.77 & 1971 & 2 &  8.49 &  0.11 \\ 
J085707.94+321032.0 & 4.776 & 4.801 & 46.94 & 46.72 & 2 & 47.25 & 2851 & 2 &  9.10 & -0.03 \\ 
J092303.53+024739.5 & 4.660 & 4.659 & 46.33 & 46.14 & 1 & 46.67 & 2636 & 1 &  8.68 & -0.18 \\ 
J093508.50+080114.5 & 4.699 & 4.671 & 46.62 & 46.33 & 2 & 46.87 & 2714 & 2 &  8.82 & -0.13 \\ 
J093523.32+411518.7 & 4.836 & 4.802 & 46.66 & 46.58 & 2 & 47.12 & 3447 & 3 &  9.18 & -0.24 \\ 
J094409.52+100656.7 & 4.748 & 4.771 & 46.64 & 46.40 & 2 & 46.93 & 2128 & 2 &  8.65 &  0.11 \\ 
J101759.64+032740.0 & 4.917 & 4.943 & 46.27 & 46.10 & 1 & 46.63 & 2822 & 1 &  8.71 & -0.26 \\ 
J105919.22+023428.8 & 4.735 & 4.789 & 46.65 & 46.36 & 2 & 46.89 & 2899 & 2 &  8.89 & -0.18 \\ 
J111358.32+025333.6 & 4.882 & 4.870 & 46.49 & 46.35 & 2 & 46.89 & 3793 & 3 &  9.12 & -0.41 \\ 
J114448.54+055709.8 & 4.793 & 4.790 & 46.13 & 46.11 & 3 & 46.63 & 3215 & 2 &  8.83 & -0.37 \\ 
J115158.25+030341.7 & 4.698 & 4.687 & 46.05 & 45.91 & 3 & 46.44 & 3741 & 3 &  8.84 & -0.57 \\ 
J120256.44+072038.9 & 4.785 & 4.810 & 46.28 & 46.27 & 1 & 46.80 & 2171 & 1 &  8.59 &  0.04 \\ 
J123503.04-000331.6 & 4.723 & 4.700 & 46.07 & 46.12 & 2 & 46.65 & 4422 & 3 &  9.11 & -0.64 \\ 
J130619.38+023658.9 & 4.852 & 4.860 & 46.57 & 46.82 & 1 & 47.35 & 5340 & 1 &  9.71 & -0.54 \\ 
J131737.28+110533.1 & 4.810 & 4.744 & 46.39 & 46.34 & 2 & 46.87 & 3144 & 2 &  8.95 & -0.25 \\ 
J132110.82+003821.7 & 4.716 & 4.726 & 46.47 & 46.17 & 2 & 46.70 & 3651 & 3 &  8.98 & -0.45 \\ 
J132853.67-022441.7 & 4.695 & 4.658 & 46.42 & 46.28 & 2 & 46.81 & 3815 & 2 &  9.08 & -0.45 \\ 
J133125.57+025535.6 & 4.737 & 4.762 & 46.15 & 46.02 & 3 & 46.55 & 3445 & 3 &  8.83 & -0.46 \\ 
J134134.20+014157.8 & 4.670 & 4.689 & 46.87 & 46.73 & 2 & 47.26 & 6480 & 2 &  9.82 & -0.74 \\ 
J134546.97-015940.3 & 4.714 & 4.728 & 46.62 & 46.08 & 2 & 46.60 & 3412 & 2 &  8.86 & -0.43 \\ 
J140404.64+031404.0 & 4.870 & 4.903 & 46.55 & 46.49 & 2 & 47.02 & 5360 & 2 &  9.51 & -0.66 \\ 
J143352.21+022714.1 & 4.721 & 4.722 & 47.14 & 46.84 & 2 & 47.37 & 2622 & 2 &  9.11 &  0.09 \\ 
J143629.94+063508.0 & 4.850 & 4.817 & 46.62 & 46.44 & 2 & 46.98 & 3052 & 2 &  8.99 & -0.19 \\ 
J144352.95+060533.1 & 4.879 & 4.884 & 46.44 & 46.16 & 3 & 46.69 & 3609 & 3 &  8.96 & -0.45 \\ 
J144734.10+102513.2 & 4.686 & 4.679 & 46.29 & 45.99 & 1 & 46.51 & 1407 & 1 &  8.03 &  0.30 \\ 
J151155.98+040803.0 & 4.686 & 4.670 & 46.62 & 46.32 & 2 & 46.86 & 1735 & 2 &  8.42 &  0.26 \\ 
J161622.11+050127.7 & 4.872 & 4.869 & 47.08 & 46.80 & 2 & 47.33 & 3910 & 1 &  9.43 & -0.27 \\ 
J165436.86+222733.7 & 4.678 & 4.717 & 47.14 & 46.48 & 1 & 47.02 & 5637 & 1 &  9.55 & -0.70 \\ 
J205724.15-003018.0 & 4.663 & 4.680 & 47.04 & 46.83 & 2 & 47.36 & 3050 & 2 &  9.23 & -0.05 \\ 
J220008.66+001744.8 & 4.818 & 4.804 & 46.70 & 46.51 & 2 & 47.04 & 2404 & 2 &  8.82 &  0.04 \\ 
J221705.72-001307.7 & 4.689 & 4.676 & 46.44 & 46.28 & 3 & 46.81 & 2274 & 3 &  8.63 &  0.00 \\ 
J222509.16-001406.8 & 4.888 & 4.890 & 46.97 & 46.70 & 1 & 47.23 & 3504 & 1 &  9.27 & -0.21 \\ 
J224453.06+134631.8 & 4.657 & 4.656 & 46.30 & 46.06 & 2 & 46.58 & 2516 & 2 &  8.58 & -0.17 
\enddata
\tablenotetext{a}{Redshift measured from the best-fit models of the \mgii\ lines.}
\tablenotetext{b}{Monochromatic luminosity at rest-wavelength 1450\AA, obtained from the SDSS/DR6 spectra and redshifts.}
\tablenotetext{c}{Quality flag associated with \Lthree. 
Quality flags of ``1'', ``2'' and ``3'' correspond to calibration and/or continuum measurement uncertainties of $\sim10\%$, $\sim20\%$ and $\sim40\%$, respectively.}
\tablenotetext{d}{Quality flag associated with \fwmg. 
Quality flags of ``1'', ``2'' and ``3'' correspond to uncertainties of $\sim10\%$, $20-30\%$ and $\sim50\%$, respectively.}
\end{deluxetable*}

To estimate \lbol, we have to 
assign a bolometric correction factor, $\fbolwv=\lbol/\lambda L_{\lambda}$.
For this we adopt the Marconi \et (2004) luminosity-dependent SED, which also  
provides a polynomial prescription for estimating $\fbol\left(4400\AA\right)$. 
The prescription can be used with other UV-optical monochromatic luminosities, by adopting the Vanden Berk \et (2001) UV continuum ($f_{\nu}\propto \nu^{-0.44}$).  
Our procedure relies on our own empirically-calibrated $\fbol\left(3000{\rm \AA}\right)$ vs. \Lthree\ relation, derived 
using the $0.5\ltsim z\ltsim0.75$ type-I SDSS AGN test sample. 
For each source, we measured both \Lop\ and \Lthree\ and converted \Lop\ to \lbol\ using the relation of Marconi \et (2004) and the Vanden Berk \et (2001) template. 
This provides, for each of the $\sim5,000$ sources, both \Lthree\ and $\fboluv=\lbol/\Lthree$. 
The best-fit relation is, 
\begin{equation}
\label{eq:fbol_uv_poly}
\fboluv= -0.58{\cal L}_{3000,44}^3 + 3.85 {\cal L}_{3000,44}^2 -8.38 {\cal L}_{3000,44} +9.34 \,,
\end{equation}
where ${\cal L}_{3000,44}\equiv \log\left(\Lthree/10^{44}\,\ergs\right)$. 
The bolometric corrections for our \zfpe\ sample range between 3.35 for the least luminous source and 3.43 for the most luminous one.
The typical \fboluv\ is lower than those used in other studies (e.g., Elvis \et 1994; Richards \et 2006b) by a factor of about 1.5.
This is mostly due to the fact that the Marconi \et (2004) estimates of \lbol\ do not include most of the mid- to far-IR emission, which originates from dust around the central source.
The normalized accretion rate is $\lledd=\lbol/\left(1.5\times10^{38}\,\mbh/\Msun\right)$ 
that is appropriate for ionized gas with solar metallicity. 
The deduced \mbh\ and \lledd\ are given in Table~\ref{tab:line_pars}. 
We have also estimated \mbh\ from the \CIV\ line, that is observed in the SDSS spectrum of each source. The line fitting procedure was
 similar to the one described above for \mgii\ and also to those discussed in other studies (e.g., Shen \et 2008; Fine \et 2010). 
We calculated \mbh\ from \Luv\ and \fwciv\ following the prescription of Vestergaard \& Peterson (2006).
We find that the \civ\ line is systematically broader than \mgii, similarly to the recent finding of Fine \et (2010). 
The differences in widths translate to higher \civ-based \mbh\ estimates, with respect to those derived from \mgii. 
These findings confirm the suspicion about the problematic use of the \civ-based method (see \S\ref{sec:intro}). 
A full analysis of these issues is deferred to a future publication.
In what follows, we rely solely on the \mgii-based measurements of \mbh\ and \lledd.

\section{Results}
\label{sec:results}

\subsection{\mbh\ and \lledd\ distributions}
\label{sec_sub:dist_mbh_lledd}

\begin{figure}
\includegraphics[width=8.5cm]{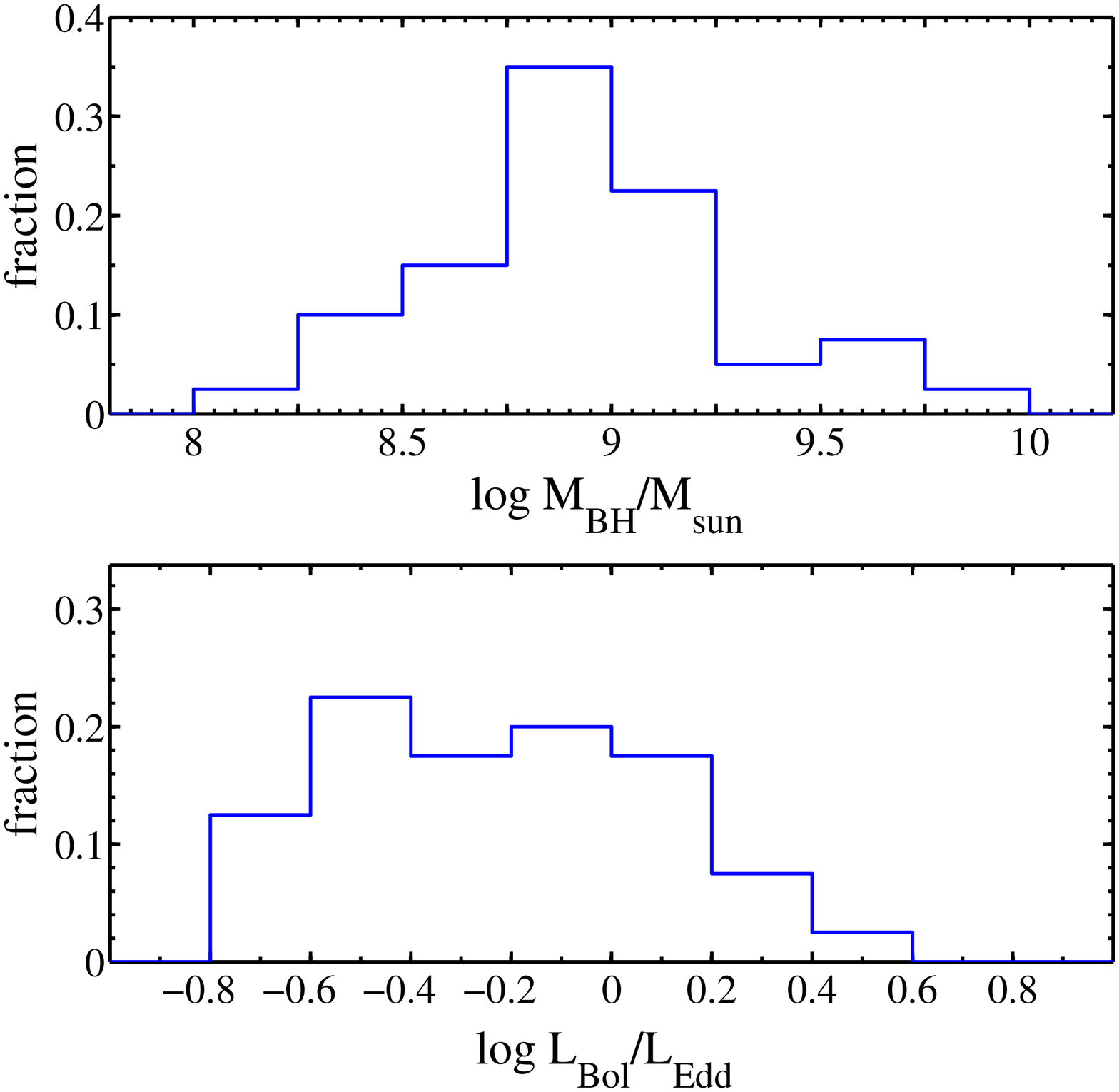} 
\caption{Distributions of \mbh\ (top) and \lledd\ (bottom) for the \zfpe\ sample.
}
\label{fig:dists_mbh_lledd}
\end{figure}

The distributions of \mbh\ and \lledd\ for the \zfpe\ sample are shown in Figure~\ref{fig:dists_mbh_lledd}. 
They cover the range 
$ 8.03 < \lmbh < 9.82$ and 
$ 0.18 < \lledd < 3.92 $. 
The median values and 68\%-percentiles (taken to be symmetric around the medians) correspond to 
$\left< \lmbh \right> = 8.89$ \raisebox{-.5ex}{$\;\stackrel{+0.32}{-0.34}\;$} and 
$\left< \lledd \right> = 0.59$ \raisebox{-.5ex}{$\;\stackrel{+0.63}{-0.30}\;$}.

As explained in \S\ref{sec:fit_mbh_lledd}, both \mbh\ and \lledd\ are derived directly from \Lthree, 
and both distributions are affected by the flux limit of the sample. 
The 1450\AA\ flux limit corresponds to $f_{\lambda}(3000\AA) \gtsim 2.1\times 10^{-18}\, \ergcmsA$ which, 
at \zfpe, gives a limiting luminosity of $\Lthree \simeq 10^{46}\, \ergs$.
The limiting \mbh\ is estimated by combining the limiting \Lthree\ and ${\rm FWHM}=1,500\kms$ using Eq.~\ref{eq:Mbh_MD04}.
This line width is the minimal width of \textit{any} emission line for which the SDSS pipeline labels the target as a spectroscopic target of type ``QSO''; a pre-requirement of our sample selection.
We thus obtain $\lmbh_{\rm limit} \simeq 8$.
Indeed, one of our sources (J1447+1025) has very similar properties, in terms of \Lthree, \fwmg\ and \mbh. 
However, this source looks as an outlier in the \mbh\ distribution. 
The next smallest source has  
$\lmbh \simeq 8.4$, which represents the low end of the \mbh\ distribution much better. 
We conclude that our sample is complete down to $\lmbh \simeq 8$, and the fact that we observe (almost) no sources with $\lmbh < 8.5$ is a real characteristic of the population of spectroscopically observed, optically selected type-I AGNs at \zfpe.

The limiting \lledd\ is deduced by combining the lowest \Lthree\ with $\fwmg=10,000\,\kms$,  
the upper boundary that FWHM(\mgii) can take in our line-fitting procedure (see \S\S\ref{sec_sub:fit}). 
The resulting limiting value is $\left(\lledd\right)_{\rm limit} \simeq 0.04$.
Clearly, the limiting \lledd\ is a factor of $\sim4.5$ below the lowest \lledd\ we actually observe.
This reflects the fact that the broadest \mgii\ line we observe has $\fwmg \simeq6,500\,\kms$.
The highest observable \lledd\ is determined by a combination of the narrowest observable \fwmg\ and the highest luminosities. 
For instance, the faint and narrow-lined source J1447+1025 is one of the higher \lledd\ sources in our sample.
On the other hand, the \textit{hightest} \lledd\ source (J0800+3051) has a similarly narrow \mgii\ line but is among the most luminous sources in the sample (a factor of $\sim5.6$ more luminous than J1447+1025), thus approaching $\lledd\simeq4$.
Since there are no brighter \zfpe\ sources in the entire SDSS/DR6 database, we conclude that our sample unveils the highest-\lledd\ ($\simeq4$) sources at \zfpe.
We also note that $\sim1/4$ of the AGNs in our sample have $1<\lledd\ltsim4$.
Several studies (e.g., Mineshige \et 2000; Wang \& Netzer 2003; Kurosawa \& Proga 2009) 
have shown that such high \lledd\ can be produced in several types of accretion disks.

\subsection{The growth times of SMBHs from $z=20$ to \zfpe}
\label{sec_sub:lifetimes}

Given the observed \mbh, \lbol\ and \lledd, we can constrain the lifetimes of the SMBHs in our sample.
We first assume that accretion proceeds with a constant \lledd. This results in an exponential growth of \mbh,  with an $e$-folding time (``Salpeter time'') of
\begin{equation}
\tau = 4 \times 10^8 \frac {\eta /(1- \eta) }{\lledd} \,\, {\rm yr} , 
\label{eq:tau_salpeter}
\end{equation}
where $\eta$ is the radiative efficiency ($\lbol=\eta\dot{M}_{\rm infall} c^2$).
The time required to grow by such accretion from a seed BH of mass \mseed\ to \mbh\ is
\begin{equation}
  t_{\rm grow} = \tau \, \ln
  \left( \frac{\mbh}{\mseed} \right) \,\, {\rm yr} .
\label{eq:t_grow}
\end{equation}
Despite several attempts to estimate $\eta$ 
(e.g., Volonteri \et 2005; King, Pringle \& Hofmann 2008, and references therein), its typical value and redshift dependence are poorly constrained. 
A broad range of $\eta\simeq0.05-0.3$ is required to match the local BH mass function to the integrated accretion history of SMBHs (e.g., Shankar \et 2004; Wang \et 2009; Shankar \et 2010a).
There is an even broader range of possible values of \mseed, depending on the various mechanisms to produce such objects. 
Remnants of population-III stars would result in $\mseed\sim10-100\,\Msun$ (e.g. Heger 
\& Woosley 2002), while collapse models of either dense stellar clusters or gas halos predict seed masses as large as $\mseed\sim10^3-10^6\,\Msun$ (e.g., Begelman, Volonteri \& Rees, 2006; Devecchi \& Volonteri 2009; see Volonteri 2010 for a review).
For the sake of consistency with earlier $z>2$ studies (N07), we assume here $\eta=0.1$ and $\mseed=10^4\,\Msun$.
With this choice of $\eta$, the $e$-folding times in our sample range from $\sim10$ to $\sim240\ {\rm Myr}$  
and the faster-accreting 50\% of the sources show $\tau \ltsim 75\ {\rm Myr}$.
Such SMBHs can increase their mass by as much as a factor of $\sim790$ within $\sim500\, {\rm Myr}$. 
We compared the growth of the SMBHs in our sample up to \zfpe\ with the corresponding age of the Universe at that redshift ($t_{\rm Universe}\simeq1.21\, {\rm Gyr}$).
The assumption of continuous, constant \lledd\ growth for such a long period is, of course, somewhat simplistic, given the timescales of standard accretion disks (a review of this and related issues can be found in King 2008).  
About 65\% of the sources have $t_{\rm growth} < t_{\rm Universe}$.
This is a larger fraction than observed in lower-redshift samples, which suggests that \zfpe\ might represent the sought after episode of fast growth for most SMBHs.

\begin{figure}
\includegraphics[width=8.5cm]{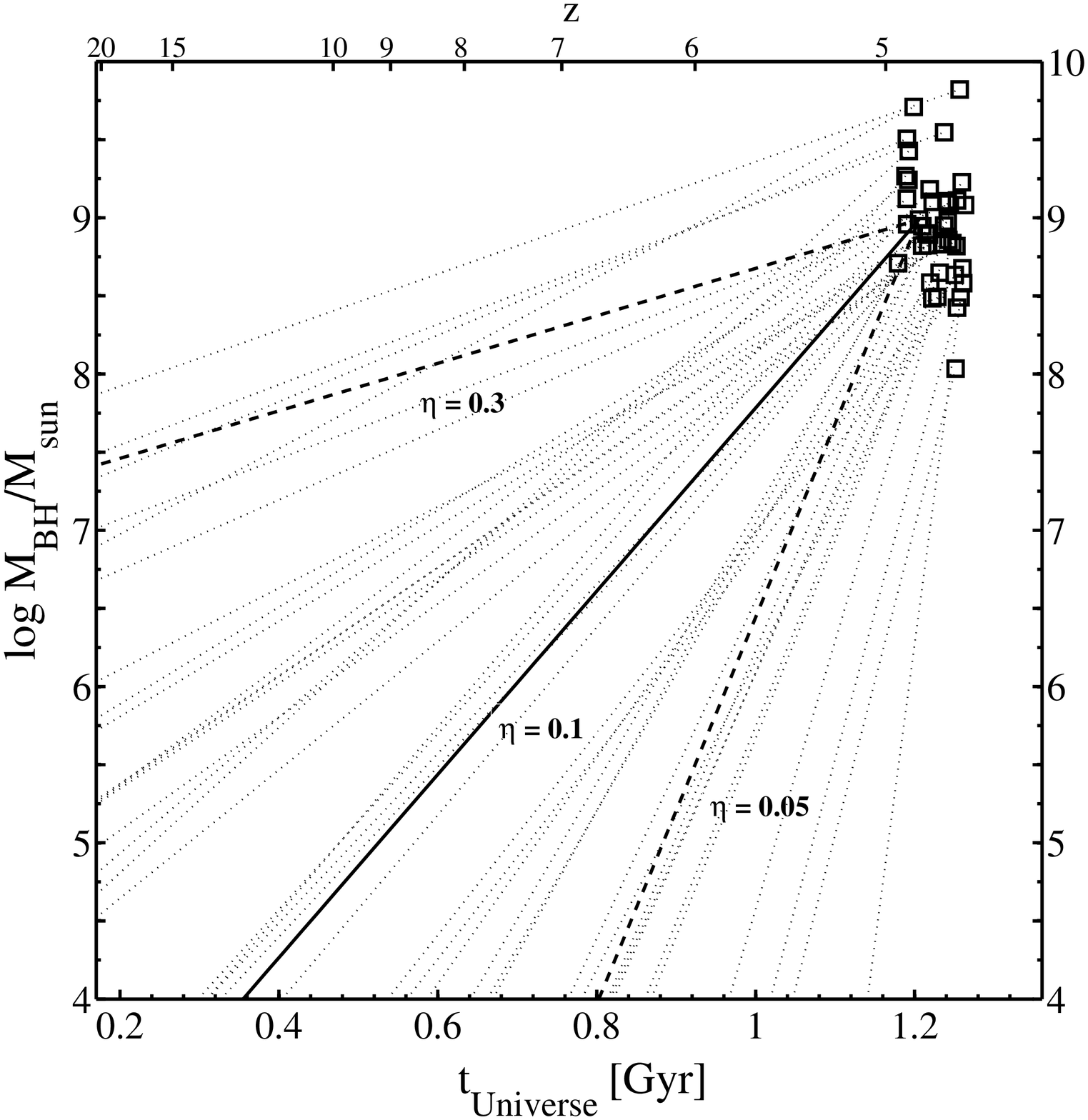}
\caption{The calculated growth of \mbh\ from $z=20$ to \zfpe. 
Black squares represent the masses of the \zfpe\ sources and the dotted lines show the regress in \mbh, assuming exponential growth with the observed \lledd\ and $\eta=0.1$.
The thick lines illustrate a choice of different values of $\eta$ for a particular source (J1436+0635) which has \mbh\ and \lledd\ that resemble the sample's median values.
}
\label{fig:m_growth_back}
\end{figure}

We also calculated the regress of \mbh\ with increasing redshift assuming the aforementioned exponential growth scenario, up to $z=20$.
These evolutionary tracks are shown in Figure \ref{fig:m_growth_back}.
The point at which such tracks cross the y-axis in Fig~\ref{fig:m_growth_back} can be regarded as the value of \mseed\ required to match the observed \mbh\ at \zfpe, given the assumption of a constant \lledd\ growth. 
The diagram demonstrates how the \zfpe\ sample can constrain the different BH formation and growth scenarios.
For example, about 40\% of the \zfpe\ sources could have grown from seed BHs with $\mseed<100\,\Msun$,  i.e. stellar-remnant BHs.
A further $\sim20\%$ could have grown from BH seeds with $1,000 < \mseed < 10^{5}\,\Msun$, and another $\sim20\%$ from BH seeds with $10^{5} < \mseed < 10^{7}\,\Msun$.
The remaining, lowest-\lledd\ sources ($\sim10\%$) have $10^{7} < \mseed < 10^{8}\,\Msun$
\footnote{Alternatively, assuming $\mseed=10^4\,\Msun$, these SMBHs have growth times which are about twice the age of the Universe at the corresponding redshift.}. 
Since even the most extreme scenarios cannot produce such high \mseed\ (Volonteri 2010), these sources had to accrete at higher rates in the past in order to attain their measured mass. 
Another explanation is a combination of massive seed BHs with low spins (i.e. small $\eta$) at very early epochs.
For example, if we assume $\eta = 0.05$ then at $z=20$ \textit{all} the sources have an implied $\mseed < 10^{6}\,\Msun$.
As mentioned above, the radiative efficiency could be higher, e.g. $\sim0.2-0.3$. 
Assuming such high values of $\eta$ in our calculations naturally leads to the requirement of more massive seed BHs. 
For example, in the extreme case of $\eta=0.3$ in all sources, we find that 95\% of the seed BHs would have $\mseed > 500\,\Msun$.

\section{Discussion}
\label{sec:discussion}

\subsection{Comparison with other studies}
\label{sec_sub:dis_compare}

Our main goal is to identify the epoch at which most SMBHs experienced the first episode of fast growth.
We focus on a comparison of our \zfpe\ sample to several $z>2$ samples with reliably measured \mbh\ and \lledd.
In S04 and N07 we studied a sample of 44 \znetprev\ type-I AGNs using NIR spectroscopy to measure 
\Lop\ and FWHM(\hb).
The study showed that a large fraction of the high \mbh\ sources accrete at a rate that is well below the Eddington limit. Assuming these sources accrete at constant \lledd, such accretion rates cannot explain their measured masses and are in contradiction with several theoretical predictions. Having obtained new data on the \zfpe\ sample, we can now compare several groups of AGNs, at several epochs, in an attempt to follow their growth all the way from \zsix\ to \ztpf.
The samples we consider here are the new one at \zfpe\ and our earlier (S04 \& N07) samples at \znetprev. We also consider a small number of sources at \zsix\  
from the samples studied by K07 and W10, which have 5 and 9 reliable \mbh\ \& \lledd\ measurements, respectively.
These small samples do \textit{not} have a common flux limit 
and simply represent the up-to-date collection of the sources discovered at those redshifts that were also observed in follow-up NIR spectroscopy.
As such, they are not representative of the AGN population at \zsix.  
This, along with the small size of the samples, limits their statistical usefulness.
To use the data from the \zsix\ samples, we re-calculated \mbh\ and \lbol\ using the methods described above.
First, we corrected the published FWHM(\mgii) according to the assumption of only one of the doublet components.
This reduces \fwmg\ by a mean factor of $\sim1.2$ 
and the reported \mbh\ by a mean factor of $\sim1.44$. 
Second, the smaller \fboluv\ adopted here reduces the reported \lbol\ by a factor of $\sim1.49$.
Regarding \lledd, the two corrections almost completely cancel out. 

\begin{figure}
\includegraphics[width=8.5cm]{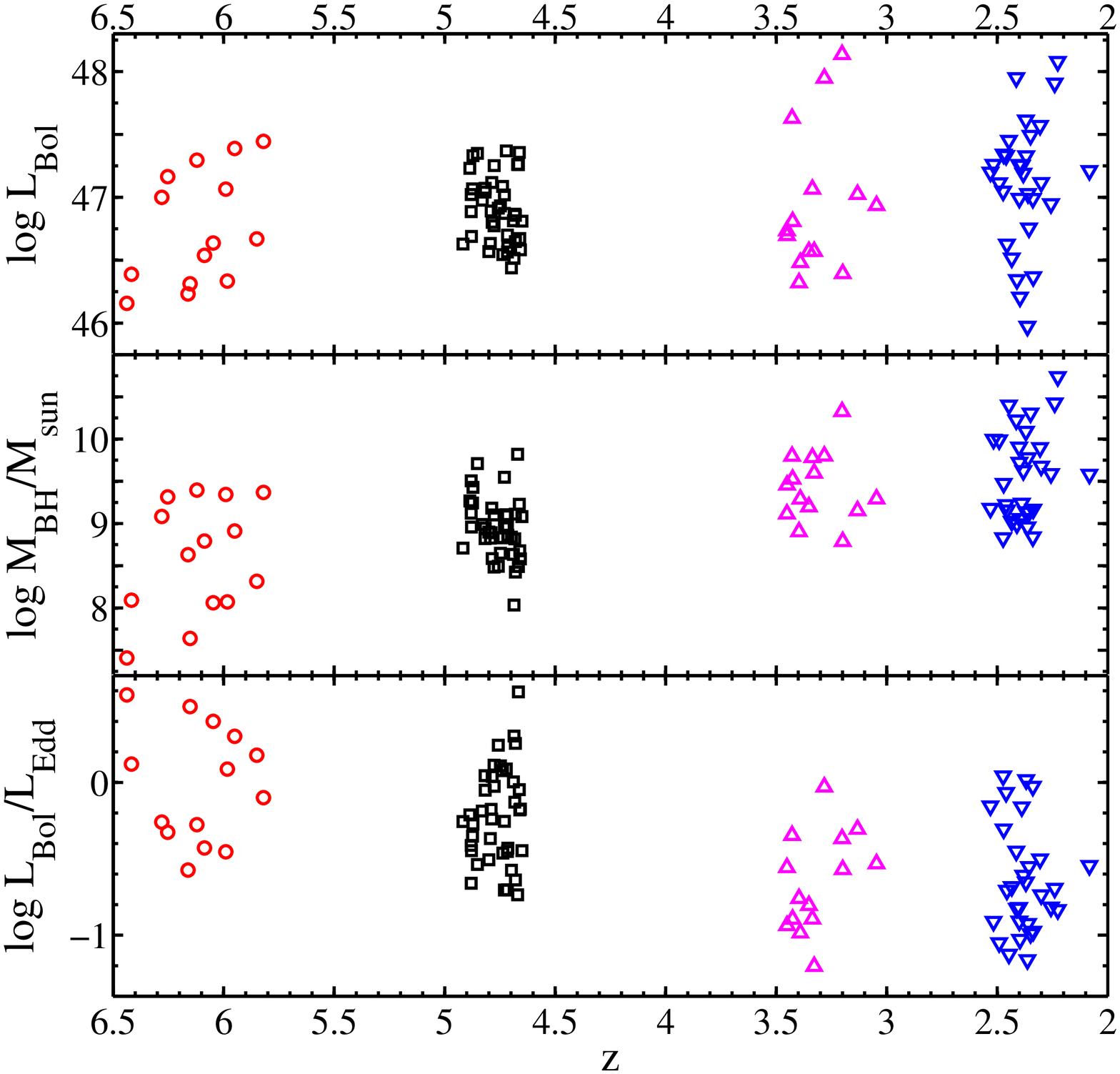}
\caption{\lbol, \mbh\ and \lledd\ vs. redshift, for samples of different redshifts discussed in the text: the new \zfpe\ sample presented here (black squares), the \znetprev\ samples of S04 and N07 (magenta and blue triangles) and the combined \zsix\ sample from K07 and W10 (red circles).}
\label{fig:all_z_props}
\end{figure}

\begin{figure}
\includegraphics[width=8.5cm]{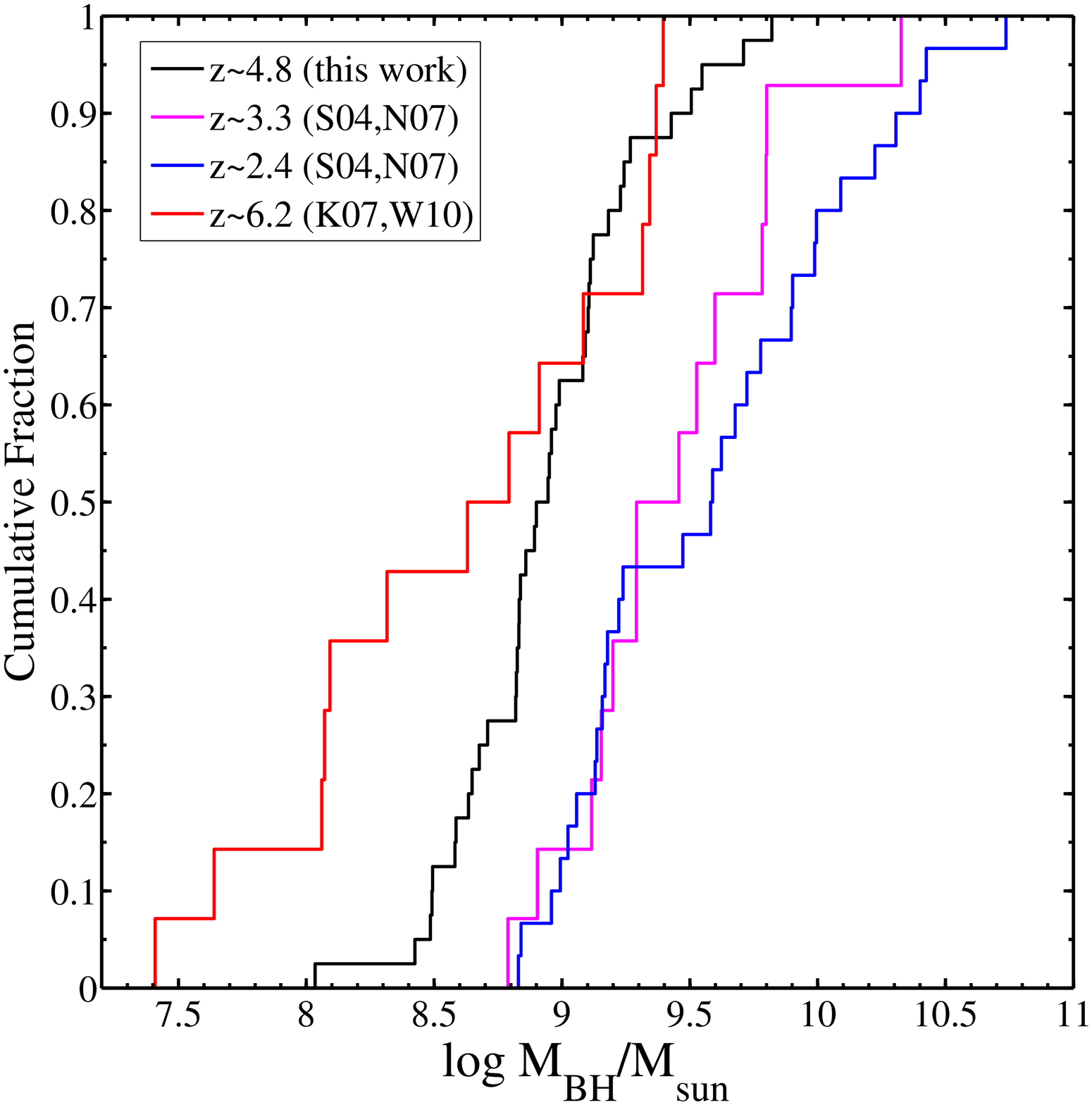}
\includegraphics[width=8.5cm]{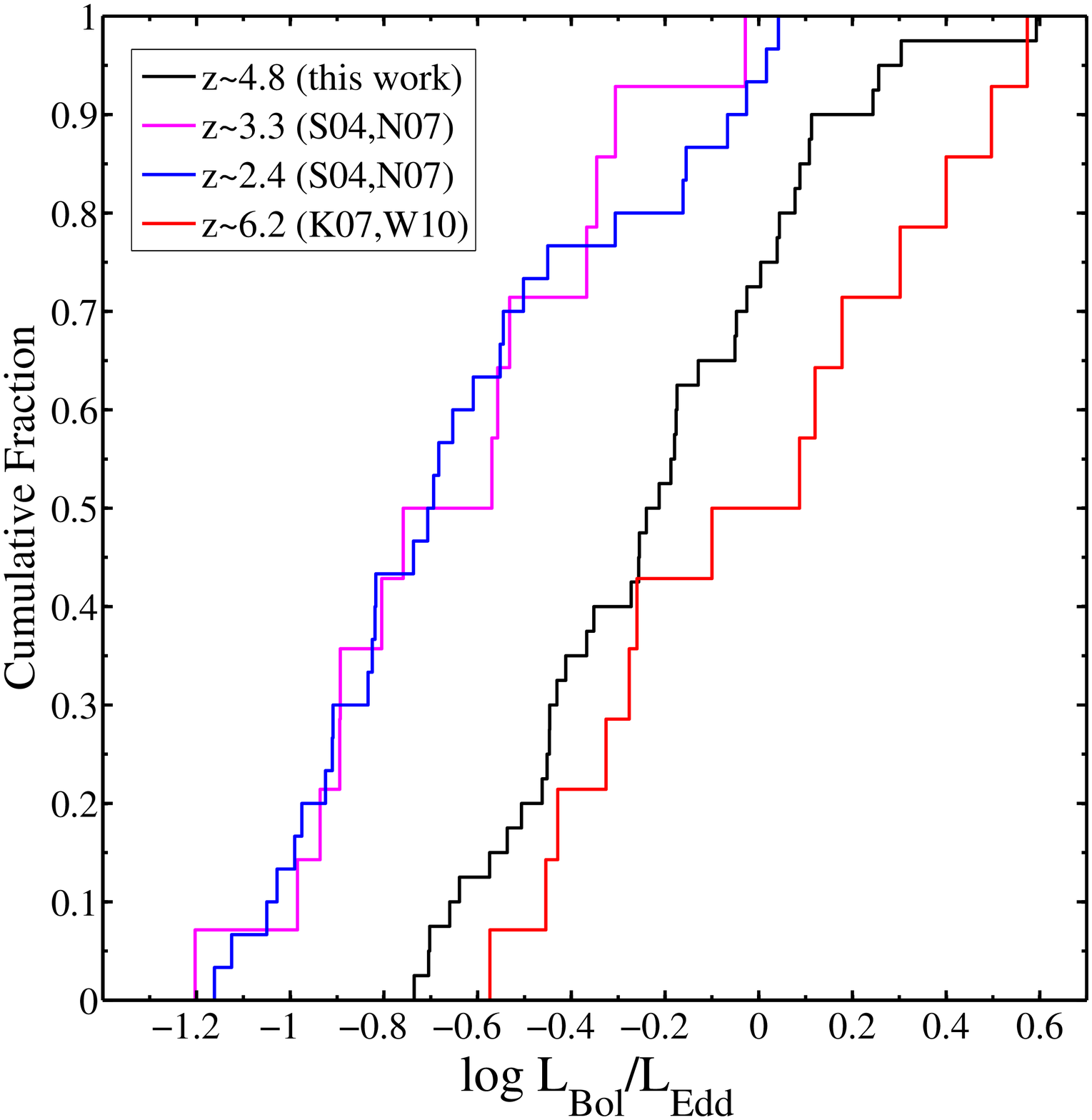}
\caption{The cumulative distribution functions (CDFs) of \mbh\ (top) and \lledd\ (bottom) for the samples discussed in the text.}
\label{fig:all_z_mbh_lledd_cdf}
\end{figure}

In Figure~\ref{fig:all_z_props} we present \lbol, \mbh\ and \lledd\ for the new \zfpe\ sample, the \znetprev\ samples of S04 and N07, and the combined sample of \zsix\ AGNs.
The four samples suggest an increasing \mbh\ and decreasing \lledd\ with cosmic time. 
This is better illustrated in Figure \ref{fig:all_z_mbh_lledd_cdf}, where we present the cumulative distribution functions (CDFs) of \mbh\ and \lledd.
The increase in the median \mbh\ and the decrease in the median \lledd\
as a function of redshift are evident. 
In particular, there is a clear shift of $\sim0.5$ dex between the median \mbh\ values of the \zfpe\ and \ztpt\ samples, in the sense of a lower \mbh\ at \zfpe.
There is also an opposite shift between the median \lledd\ values.
The differences between the \ztpt\ and the \ztpf\ samples are much smaller, although the \ztpf\ sample includes several sources with extremely massive BHs ($\lmbh>10$) which are not observed at earlier epochs.
Regarding \mbh,  
only $\sim14\%$ of the \ztpt\ sample (2 sources) lie below the median value of the \zfpe\ sample ($\lmbh=8.92$).
Similarly, only $\sim13\%$ of the \zfpe\ sample (5 sources) lie above the median value of the \ztpt\ sample ($\lmbh=9.37$) and even less ($\sim6\%$; 2 sources) above the median value of the \ztpf\ sample ($\lmbh=9.59$).
%

%

%

%
We have further tested the significance of these differences by performing a series of two-sample Kolmogorov-Smirnov tests.
The null hypothesis that the observed distributions of \lledd\ at \zfpe\ and at \ztpt\ (or at \ztpf) are drawn from the same parent distribution is rejected with significance levels $>99\%$.
A similar test was applied to the \zfpe\ and \zsix\ samples, and could not reject the null hypothesis, suggesting that the distributions of \lledd\ at \zfpe\ and \zsix\ are statistically similar.
We obtain similar results when comparing the distribution of \mbh\ at \zfpe\ to that of the \znetprev\ samples.
Due to the large fraction of \zsix\ sources with $\lmbh\ltsim8.5$, we can reject the hypothesis that the \mbh\ values at \zfpe\ and \zsix\ are drawn from the same distribution.
Some of the above results may be biased by the fact that the four samples cover a different range of \lbol, which originate from the different target selection criteria. 
We thus repeated the statistical tests aforementioned, focusing on subsamples which share a common range of $46.4 < \llbol < 47.4$ (i.e., matching the \lbol\ range of the \zfpe\ sample). 
All but one comparison result in the same conclusions, with similar confidence levels. 
The one exception are the distributions of \mbh\ at \zfpe\ and \zsix, for which the null hypothesis now \textit{cannot} be rejected, suggesting that these distributions represent the same parent population.
This is not surprising given the lower luminosities of most of the \zsix\ sources, the dependence of \mbh\ on source luminosity, and the incompleteness of the \zsix\ sample. 
We conclude that there is strong evidence for a rise in \mbh\ and a drop in \lledd\ of about a factor of 2.8 between \zfpe\ and \ztpt. 
This strong trend is \textit{not} observed with respect to neither lower nor higher-redshift samples, which span similar periods of time
\footnote{For the adopted cosmology, the physical time between $z=4.8$ and $z=3.3$ is $\sim680\,{\rm Myr}$ and between $z=3.3$ and $z=2.4$ is $\sim790\,{\rm Myr}$.}.
It thus seems that the most massive BHs, associated with the most luminous AGNs, started the episode of fast BH growth at redshifts above about $z\sim5$. 
By $z\sim2-3$, these SMBHs reach their peak (final) mass ($\>10^{10}\,\Msun$) and their mass accretion is less efficient.

\subsection{The growth of active SMBHs from \zfpe\ to \ztpt\ and \ztpf}
\label{sec_sub:dis_z65_z2}

\begin{figure}
\includegraphics[width=8.5cm]{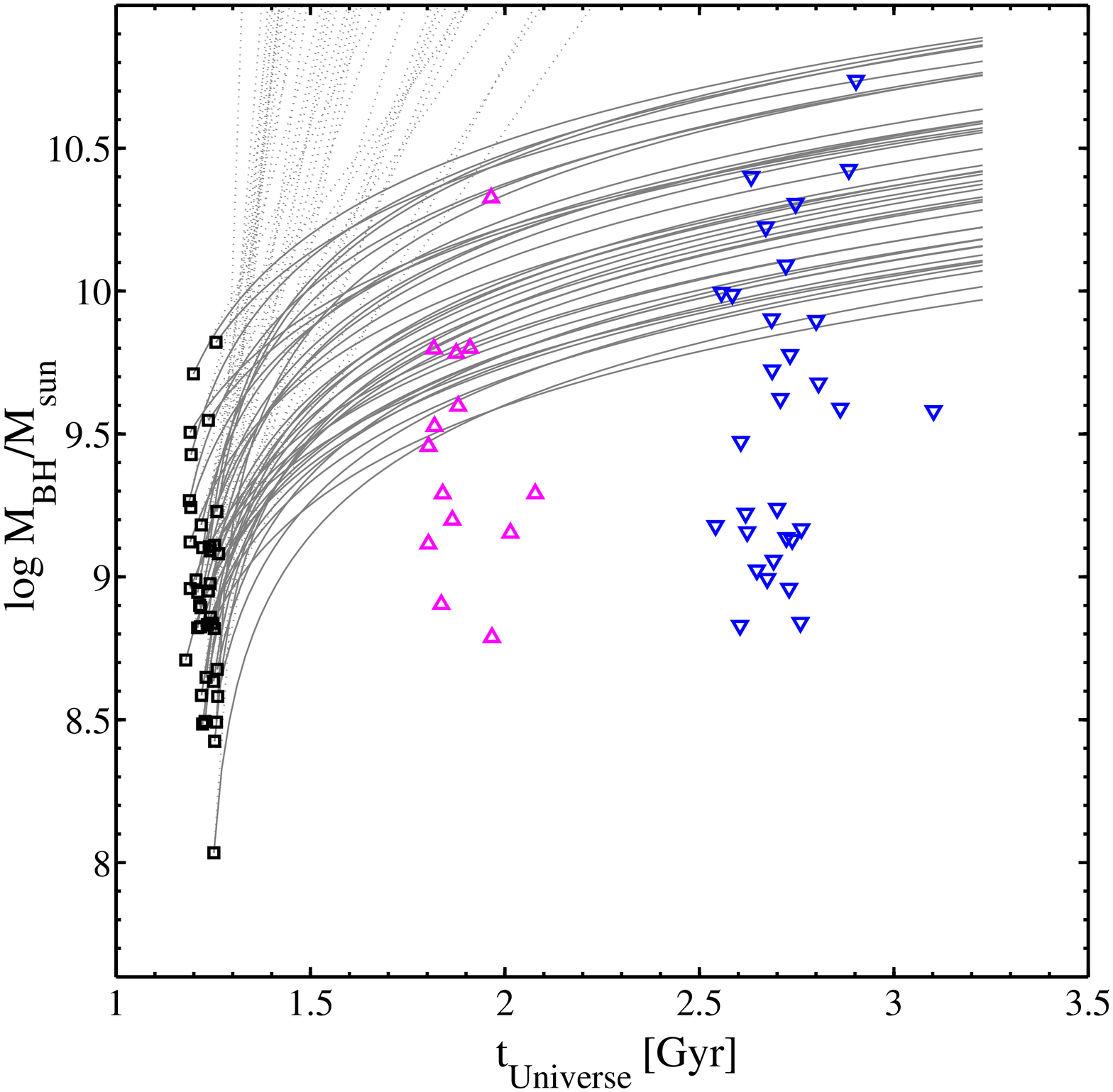}
\caption{Evolution scenarios for the \zfpe\ SMBHs. 
Symbols are identical to those in Fig.~\ref{fig:all_z_props}.
Solid lines describe the growth of \mbh\ under the assumption of a constant \lbol, while dotted lines represent the constant \lledd\ scenario.}
\label{fig:m_growth_for}
\end{figure}

Given the trends and differences in \mbh\ and \lledd, as well as the short $e$-folding times of the \zfpe\ AGNs (\S\S\ref{sec_sub:lifetimes}), 
it is possible to think of the \zfpe, \ztpt\ and \ztpf\ samples as representing different evolutionary stages of the same parent population of SMBHs. 
In what follows, we focus on the \zfpe\ and \ztpt\ samples (separated by $\sim 680\, {\rm Myr}$) to test this evolutionary interpretation, assuming various scenarios. 
We consider two growth scenarios: constant accretion rate (i.e. constant \lbol) and constant \lledd
\footnote{There are, of course, many other possible scenarios involving, for example, host related evolution. 
These are beyond the scope of the present paper.}.
As explained, the assumption of constant \lledd\ results in an exponential growth of \mbh\ and given our chosen value of $\eta$, the mass growth $e$-folding time is $\tau\simeq45\left(\lledd\right)^{-1}\,{\rm Myr}$, which translates to  $\tau\ltsim 240\,{\rm Myr}$ ($\lledd>0.18$) for the \zfpe\ sources. 
Thus, even the lowest \lledd\ SMBHs at \zfpe\ could have increased their \mbh\ by a factor of $\sim20$ by \ztpt, which is much larger than the typical difference in \mbh\ between the two samples.
The assumption of a constant \lbol\ results in much slower growth. 
The luminosities of the \zfpe\ sources translate to $4\ltsim  \dot{M}_{\rm BH} \ltsim 37 \, \mpyr$.
Even this much slower growth scenario results in very high \mbh\ at \ztpt.
This scenario also produces very low accretion rates ($\lledd<0.1$) at \ztpt.
In Figure \ref{fig:m_growth_for} we present evolutionary tracks for our \zfpe\ AGNs to \ztpt\ and eventually to $z=2$.
Clearly, continuous constant \lbol\ growth results in too large \mbh\ and cannot reproduce the lower \mbh\ sources at \znetprev. 
Specifically, the calculated distribution of \mbh\ for the \zfpe\ sources, when evolved to \ztpt, has a median value of $\lmbh\simeq10$, larger by $\sim0.6$ dex. than the \textit{observed} median of the \ztpt\ sample.
This scenario also fails to reproduce the observed range of \lledd, since the lowest \lledd\ sources at \ztpt\ have $\lledd\gtsim0.1$.
As the dotted evolutionary tracks in Fig~\ref{fig:m_growth_for} demonstrate, the constant \lledd\ scenario produces even larger masses, and is only feasible if all the \zfpe\ sources cease their accretion shortly after their observed active phase. 
This scenario assumes a constant \lledd, hence the significant difference between the \lledd\ distributions at \zfpe\ and at \ztpt\ ($\sim0.5$ dex ; see Fig~\ref{fig:all_z_mbh_lledd_cdf}) is not resolved. 
We also note that in both scenarios the \zfpe\ sources would have  
been easily observed at \ztpt, since they would have \lbol\ which is either similar (linear growth) or much higher than (exponential growth) the sources observed by S04 and N07 at \znetprev.
From these two simplistic growth scenarios we conclude that the fast growth of the \zfpe\ sources must either experience a shut down before \ztpt, or accrete in several short episodes, with duty cycles which are much smaller than unity.

\begin{figure}
\includegraphics[width=8.5cm]{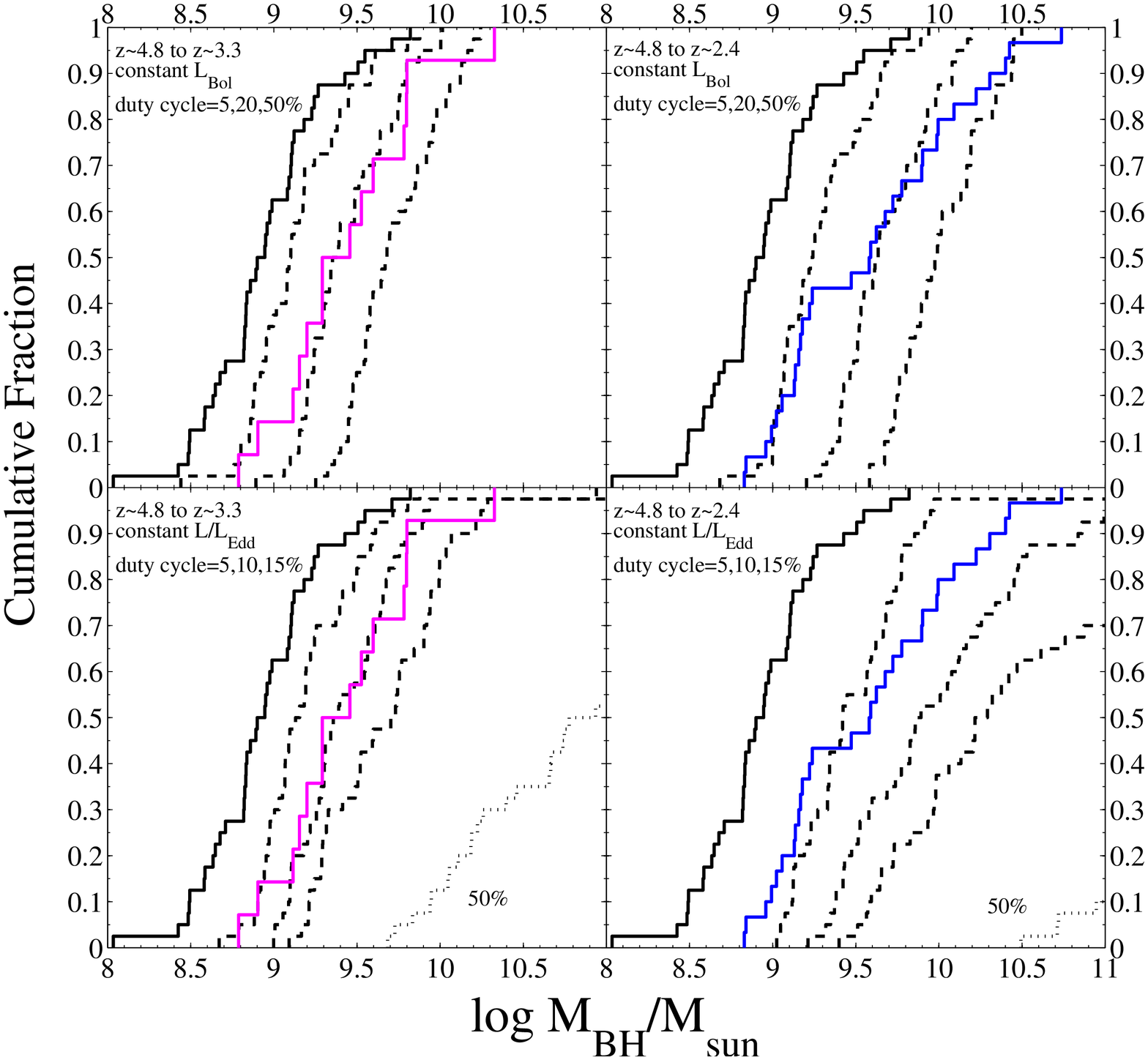}
\caption{Observed and calculated \mbh\ CDFs for different evolutionary scenarios.
In all panels solid black lines show the observed \mbh\ CDF at \zfpe, while magenta lines (left panels) and blue lines (right panels) show the CDFs at \ztpt\ and at \ztpf, respectively.
Dashed black lines are the calculated CDF of the \zfpe\ sample assuming different growth scenarios and duty cycles.  
Top panels assume a constant \lbol\ scenario, and duty cycles of 5, 20, and 50\% (from left to right).
Bottom panels assume a constant \lledd\ scenario, and duty cycles of 5, 10 and 15\% (from left to right).
The dotted lines in the bottom panels illustrate how a duty cycle of 50\% produces extremely over-massive BHs at \ztpt\ and \ztpf.
All the calculations assume $\eta=0.1$.
Assuming the extreme cases of $\eta=0.05$ (or $0.3$) would mean that the plotted calculated CDFs correspond to duty cycles which are a factor of $\sim2$ smaller (or larger; see text for details).
}
\label{fig:mbh_cdf_evolve_for}
\end{figure}

To further constrain the evolution of the observed SMBHs, we ran a series of calculations with different duty cycles for each of the two evolutionary scenarios. 
In each calculation we assembled the distribution of calculated \mbh\ at exactly $z=3.3$ and $z=2.4$.
Several of these distributions are shown in Figure \ref{fig:mbh_cdf_evolve_for}.
For the fast, constant \lledd\ scenario, the only calculated distributions which resemble the observed distribution of \mbh\ at \ztpt\ are those with a duty cycles in the range $7.5-12.5\%$. 
The observed distribution at \ztpf, on the other hand, can only be achieved by duty cycles of $5-7.5\%$.
For the constant \lbol\ scenario, the \ztpt\ distribution can be matched by assuming a duty cycle in the range of $15-25\%$, while the \ztpf\ distributions can be partially explained by assuming duty cycles of $\sim5-25\%$.
The ranges of duty cycles which account for the observed distribution of \lledd\ are somewhat different: $10-15\%$ for the \lledd\ distribution at \ztpt\ and $5-10\%$ for the one at \ztpf.
All the above calculations assumed $\eta=0.1$. 
Naturally, lower (higher) radiative efficiencies will require shorter (longer) duty cycles, to reproduce the same calculated CDFs at \znetprev. 
This means that for any assumed $\eta$ in the range $0.05\leq\eta\leq0.3$ the aforementioned ``acceptable'' duty cycles can change by up to a factor of $\sim2$, where the exact factor scales as $\eta /(1- \eta)$.
We note that Fig.~\ref{fig:mbh_cdf_evolve_for} suggests that, in both evolution scenarios, the more massive BHs at \znetprev\ seem to have grown at higher duty cycles than the less massive ones.
Alternatively, this can be interpreted as an increase in radiative efficiency with increasing resultant \mbh\ at \znetprev.  
In particular, the observed distributions of \mbh\ and \lledd\ at \ztpf\ are much more complex than the calculated ones and no single, fixed duty cycle can account for the shape of the observed distributions.
The reason for this apparent discrepancy might also be the way the \ztpf\ sources were selected in the S04 and N07 studies.

\begin{figure}
\includegraphics[width=8.5cm]{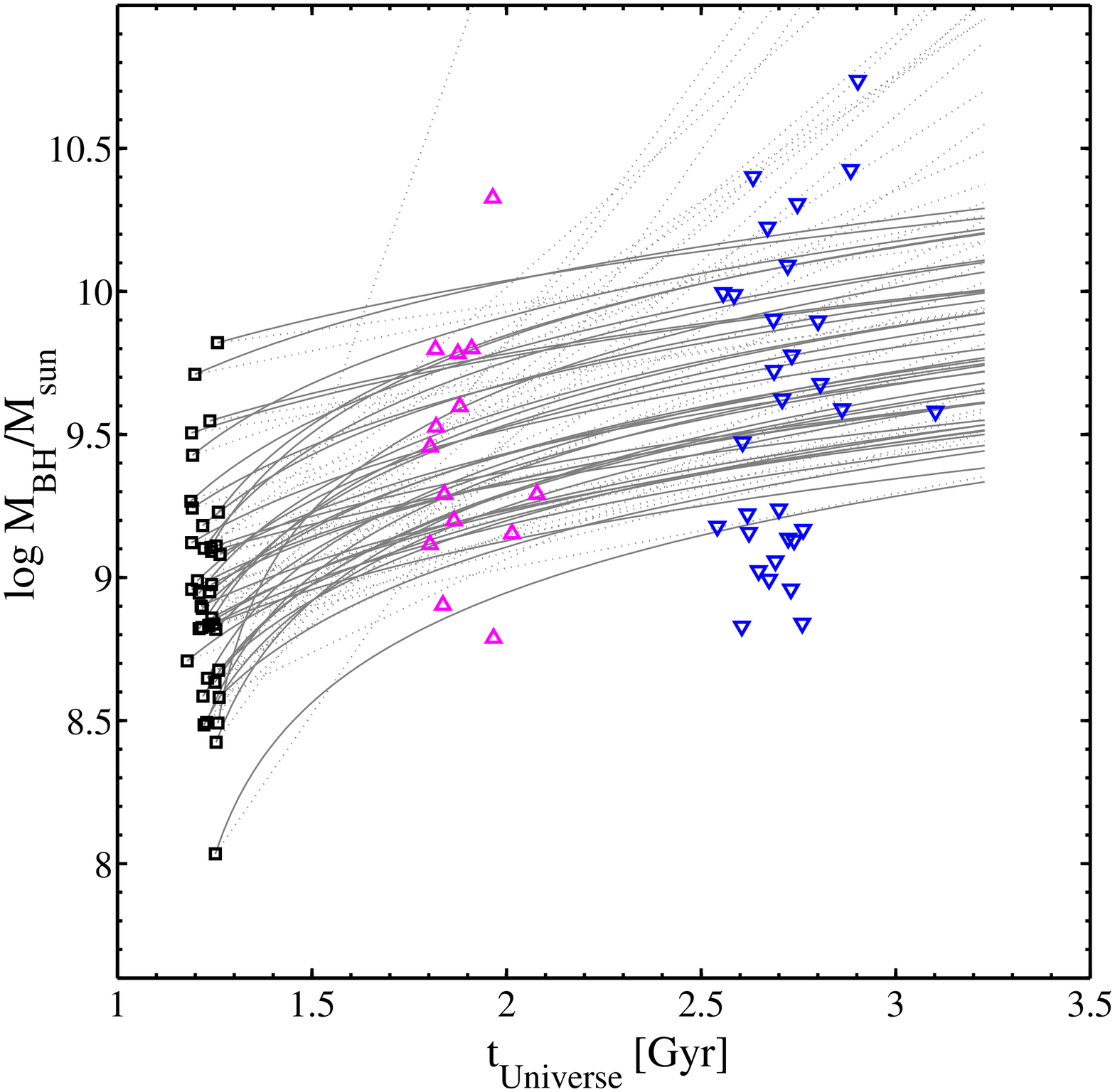}
\caption{Evolution scenarios for the \zfpe\ SMBHs with various duty cycles.
Symbols are identical to those in Fig.~\ref{fig:all_z_props}. 
Solid lines describe \mbh\ growth under the assumption of constant \lbol\ and a duty cycle of 20\%. 
Dotted lines represent the constant \lledd\ scenario and a duty cycle of 10\%.}
\label{fig:mbh_t_duty_cycle}
\end{figure}

To illustrate how the above duty cycles facilitate an evolutionary connection between the three samples, we present in Figure \ref{fig:mbh_t_duty_cycle} the evolutionary tracks of the \zfpe\ sample, similar to Fig~\ref{fig:m_growth_for}, but this time assuming duty cycles of 10\% and 20\%, for the constant \lledd\ and constant \lbol\ scenarios, respectively and assuming again $\eta=0.1$.
We also verified that evolving the \mbh\ of the \znetprev\ sources \textit{backwards}, under the assumption of constant \lledd\ and a duty cycle of $10\%$ results in a distribution of \mbh\ which is not critically different than the one directly observed at \zfpe. 
However, we again find that the few highest \mbh\ sources at \ztpf\ probably require higher duty cycles. 

We conclude that all the observed measurements of \mbh, \lledd\ and \lbol\ are consistent with duty cycles of about 10-20\%. 
Considerably longer duty cycles can be consistent with observations only if we assume an extremely high radiative efficiency ($\eta\simeq0.3$) for all the \zfpe\ sources.
Such low duty cycles are in good agreement with the models presented in Shankar \et (2009), which are able to reproduce the growth of a similar population of SMBHs, in terms of redshift and range of \mbh\ (c.f. their Fig.~7).. 
Our constrains on duty cycles are, however, in contrast with several other studies that suggest that the duty cycle at high redshift should reach unity, based on the clustering of high-redshift AGNs (e.g. White \et 2008; Wyithe \& Loeb 2009; Bonoli et al. 2010, and references therein).
All this leads us to suggest that for the most massive $z>2$ type-I AGNs, an epoch of fast SMBH growth took place before $z\sim4$, which we partially observe at \zfpe.
This epoch is efficient enough to produce the very massive ($\lmbh\gtsim10$) BHs observed at \ztpf.
The fast growth slows down before \ztpt, and the sources observed at \ztpt\ show much lower \lledd. 
In addition, there is no significant rise in \mbh\ between \ztpt\ and \ztpf.
The growth of \mbh\ during this epoch can only be explained by assuming that mass accretion proceeded in short episodes, lasting an order $10-20\%$ of the total period.
This translates to accretion episodes which (cumulatively) last $\sim70-140\, {\rm Myr}$ ( $\sim150-300\, {\rm Myr}$) over the period between \zfpe\ and \ztpt\ (\zfpe\ and \ztpf), respectively. 
Another possibility is that the \zfpe\ SMBHs have experienced an even faster shut down, and so at \ztpt\ their inactive relics have masses which do not differ significantly from those we observe at \zfpe.

Major mergers between massive, gas-rich galaxies are capable of supplying large amounts of cold gas directly to the innermost regions, to be accreted by the central SMBHs. 
Detailed simulations of such mergers suggest that these events can fuel significant BH growth over periods of $\sim1\, {\rm Gyr}$.  
The mass accretion rate and source luminosity during such mergers may vary on time-scales of $few \times 10\, {\rm Myr}$ (Di Matteo \et 2005; Hopkins \et 2006; Hopkins \& Hernquist 2009). 
According to Hopkins \et (2006), AGNs would appear to have $\lbol\gtsim2.7\times10^{46}\,\ergs$ (i.e., matching the range we observe at \zfpe) for typically $\ltsim100\,{\rm Myr}$.
Numerical studies suggest that massive dark matter halos may undergo more than one major mergers per ${\rm Gyr}$ at \zfpe\ (e.g. Genel \et 2009 and references therein).    
Thus, during the $\sim680\, {\rm Myr}$ between \zfpe\ and \ztpt, the hosts of the \zfpe\ SMBHs may have undergone a single merger, during which the period of significant accretion by the SMBHs would last for $\ltsim100\, {\rm Myr}$. 
This is in good agreement with the duty cycles of $10-20\%$ we find here, which correspond to $70-140\,{\rm Myr}$.
The decline in activity observed for the most massive BHs at \ztpt\ and \ztpf\ may then be associated with the decline in the major merger rate, which drops by a factor of $\sim4$ between \zfpe\ and \ztpf\ (Genel \et 2009).
However, it is likely that the hosts of the \ztpf\ SMBHs have experienced an additional major merger during the $\sim790\,{\rm Myr}$ between \ztpt\ and \ztpf, especially if these SMBHs reside in the more massive dark matter halos.
In such a scenario, the largest BHs at \ztpf\ may have gathered their high mass during more efficient (i.e., higher duty-cycle) accretion episodes.
This requirement of higher duty-cycles for higher \mbh\ sources is also reflected in our analysis (see upper-right panel of Fig.~\ref{fig:mbh_cdf_evolve_for}). 
If major mergers are indeed the main drivers of SMBH accretion history at $z\sim2-5$, the results presented here predict that the host galaxies of the \zfpe\ sources would be found in the early stages of major mergers, while the hosts of \znetprev\ sources would appear to be in either later stages, or in merging systems which have a lower mass ratio.
There are several other mechanisms to make large amounts of cold gas available for BH accretion.
Most of these might also trigger intense star formation (SF) activity in the hosts of SMBHs.
Indeed, many high-luminosity AGNs show evidence for intense SF (e.g. Netzer 2009b; Lutz \et 2010, and references therein).
New observations by \textit{Herschel} suggest that SF in the hosts of the most luminous AGN peaks at $z\sim3$ and quickly decreases at later epochs (Serjeant \et 2010). 
If correct, it will indicate that the amount of gas available for both BH accretion and SF has depleted by $z\sim2-3$, consistent with the decrease in SMBH accretion activity we find here.
Future \textit{Herschel} and ALMA observations of the hosts of the \zfpe\ AGNs may be able to reveal the presence of such SF activity, and perhaps the amount and dynamical state of the cold gas.
This will enable a better understanding of the galaxy-scale processes which drive the accretion history of the fast-growing \zfpe\ SMBHs.

\section{Summary}
\label{sec:summary}
 
We present new \hband\ spectroscopy for a flux-limited sample of \Ntot type-I SDSS AGNs at \zfpe.  
The sample covers $\sim1/4$ of all the  (spectroscopically observed) SDSS sources at that redshift band, and thus about $\sim1/20$ of the total population of $\lbol > 2.75\times10^{46} \ergs$ sources (over the entire sky).
The main results of our study are:

\begin{enumerate}
\item 
The \zfpe\ AGNs have, on average, higher accretion rates and lower massees than those observed at lower redshifts.
The accretion rates and masses are comparable to those of the small, incomplete samples of \zsix\ AGNs.

\item
We have observed an epoch of fast SMBH growth, probably the very first such phase for most SMBHs.
Assuming continuous growth from about $z=20$, these observations provide the very first look at the distribution of seed BHs in the early universe.
About 65\% of the SMBHs at \zfpe\ have had enough time to grow to their observed \mbh, assuming continuous accretion at the observed \lledd.
About $40\%$ of the sources could have started their growth from BH seeds which are stellar remnants 
($\mseed < 100\,\Msun$).
For the minority the sources, those with small \lledd,
there might have been an ever earlier epoch of faster accretion or, perhaps, they started their growth from a much larger seeds ($\mseed\gtsim10^{6}\, \Msun$). 

\item
The \zfpe\ sources can be regarded as the progenitor population of the most massive ($\mbh\gtsim10^{10}\,\Msun$) BHs observed at \znetprev. 
The growth rate of those massive BHs seems to be much slower between \ztpt\ and \ztpf. 

\item
A comparison of the observed distributions of \mbh\ at \zfpe, \ztpt\ and \ztpf\ indicates that the \zfpe\ sources either completely stop their accretion shortly after \zfpe, or that their accretion proceeds in relatively short episodes. We find that for mass growth rates that follow either constant \lledd\ or constant \lbol\ scenarios, duty cycles of either $\sim10\%$ or $\sim20\%$, respectively, give reasonable agreement to the observed distributions of \mbh.

\end{enumerate}

\acknowledgments
We thank the referee for his/her detailed comments and suggestions.
Funding for this work has been provided by the Israel Science Foundation grant 364/07
and by the Jack Adler Chair for Extragalactic Astronomy.
BT acknowledges generous support by the Dan David Foundation.
PL is grateful of support by Fondecyt project \#1080603.
This work is based on observations collected at the European Organization for Astronomical Research in the Southern Hemisphere, Chile, 
and in addition on observations obtained at the Gemini Observatory, which is operated by the
Association of Universities for Research in Astronomy, Inc., under a cooperative agreement
with the NSF on behalf of the Gemini partnership: the National Science Foundation (United
States), the Science and Technology Facilities Council (United Kingdom), the
National Research Council (Canada), CONICYT (Chile), the Australian Research Council
(Australia), Minist\'{e}rio da Ci\^{e}ncia e Tecnologia (Brazil) 
and Ministerio de Ciencia, Tecnolog\'{\i}a e Innovaci\'{o}n Productiva  (Argentina)
This work makes use of data from the SDSS.
Funding for the SDSS and SDSS-II has been provided by the Alfred P. Sloan Foundation, the Participating Institutions, the National Science Foundation, the U.S. Department of Energy, the National Aeronautics and Space Administration, the Japanese Monbukagakusho, the Max Planck Society, and the Higher Education Funding Council for England. The SDSS Web Site is http://www.sdss.org/.
%

%


%
%



\newpage


\begin{thebibliography}{}
 
\bibitem[Adelman-McCarthy et al.(2008)]{2008ApJS..175..297A} 
Adelman-McCarthy, J.~K., et al.\ 2008, \apjs, 175, 297 
 

\bibitem[Baldwin et al.(2004)]{2004ApJ...615..610B} Baldwin, J.~A., 
Ferland, G.~J., Korista, K.~T., Hamann, F., 
\& LaCluyz{\'e}, A.\ 2004, \apj, 615, 610 
 
 
 
 
\bibitem[Baskin \& Laor(2005)]{2005MNRAS.356.1029B} Baskin, A., \&
  Laor, A.\ 2005, \mnras, 356, 1029
 
 
\bibitem[Bauer et al.(2009)]{2009ApJ...696.1241B} Bauer, A., Baltay, C., 
Coppi, P., Ellman, N., Jerke, J., Rabinowitz, D., 
\& Scalzo, R.\ 2009, \apj, 696, 1241 

 
\bibitem[Becker et al.(1995)]{1995ApJ...450..559B} Becker, R.~H., White, 
R.~L., \& Helfand, D.~J.\ 1995, \apj, 450, 559 
 

 

\bibitem[Begelman et al.(2006)]{2006MNRAS.370..289B} Begelman, M.~C., 
Volonteri, M., \& Rees, M.~J.\ 2006, \mnras, 370, 289 

 

\bibitem[Bentz et al.(2009)]{2009ApJ...697..160B} Bentz, M.~C., Peterson, 
B.~M., Netzer, H., Pogge, R.~W., \& Vestergaard, M.\ 2009, \apj, 697, 160 
 

\bibitem[Bonoli et al.(2010)]{2010MNRAS.404..399B} Bonoli, S., Shankar, F., 
White, S.~D.~M., Springel, V., \& Wyithe, J.~S.~B.\ 2010, \mnras, 404, 399 



 
\bibitem[Corbett et al.(2003)]{2003MNRAS.343..705C} Corbett, E.~A., et al.\ 
2003, \mnras, 343, 705 

 
\bibitem[Croom et al.(2009)]{2009MNRAS.399.1755C} Croom, S.~M., et al.\ 
2009, \mnras, 399, 1755 

 
\bibitem[Cutri et al.(2003)]{2003tmc..book.....C} Cutri, R.~M., et al.\ 
 2003, The IRSA 2MASS All-Sky Point Source Catalog, NASA/IPAC Infrared 
 Science Archive.~
http://irsa.ipac.caltech.edu/applications/Gator/
 

\bibitem[Davies(2007)]{2007MNRAS.375.1099D} Davies, R.~I.\ 2007, \mnras, 
375, 1099 



\bibitem[Devecchi 
\& Volonteri(2009)]{2009ApJ...694..302D} Devecchi, B., \& Volonteri, M.\ 2009, \apj, 694, 302 
 

\bibitem[Di Matteo et al.(2005)]{2005Natur.433..604D} Di Matteo, T., 
Springel, V., \& Hernquist, L.\ 2005, \nat, 433, 604 
 

 
 
\bibitem[Eisenhauer et al.(2003)]{2003SPIE.4841.1548E} Eisenhauer, F., et 
al.\ 2003, \procspie, 4841, 1548 
 
 
\bibitem[Elvis et al.(1994)]{1994ApJS...95....1E} Elvis, M., et al.\ 1994, 
\apjs, 95, 1 
 

\bibitem[Fan et al.(2006)]{2006AJ....132..117F} Fan, X., et al.\ 2006, \aj, 
132, 117 
 
 

\bibitem[Fine et al.(2008)]{2008MNRAS.390.1413F} Fine, S., et al.\ 2008, 
\mnras, 390, 1413 
 


\bibitem[Fine et al.(2010)]{2010arXiv1005.5287F} Fine, S., Croom, S.~M., 
Bland-Hawthorn, J., Pimbblet, K.~A., Ross, N.~P., Schneider, D.~P., 
\& Shanks, T.\ 2010, arXiv:1005.5287 




\bibitem[Genel et al.(2009)]{2009ApJ...701.2002G} Genel, S., Genzel, R., 
Bouch{\'e}, N., Naab, T., \& Sternberg, A.\ 2009, \apj, 701, 2002 
  
 
\bibitem[Hasinger et 
al.(2005)]{2005A&A...441..417H} Hasinger, G., Miyaji, T., \& Schmidt, M.\ 2005, \aap, 441, 417 
 
 
 
 
\bibitem[Heger 
\& Woosley(2002)]{2002ApJ...567..532H} Heger, A., \& Woosley, S.~E.\ 2002, \apj, 567, 532 


 
\bibitem[Hodapp et al.(2003)]{2003PASP..115.1388H} Hodapp, K.~W., et al.\ 
2003, \pasp, 115, 1388 
 
 
 
\bibitem[Hopkins et al.(2006)]{2006ApJS..163....1H} Hopkins, P.~F.,
  Hernquist, L., Cox, T.~J., Di Matteo, T., Robertson, B., \&
  Springel, V.\ 2006, \apjs, 163, 1

 
\bibitem[Hopkins 
\& Hernquist(2009)]{2009ApJ...698.1550H} Hopkins, P.~F., \& Hernquist, L.\ 2009, \apj, 698, 1550 
 

 
\bibitem[Kaspi et al.(2000)]{2000ApJ...533..631K} Kaspi, S., Smith, P.~S., 
Netzer, H., Maoz, D., Jannuzi, B.~T., \& Giveon, U.\ 2000, \apj, 533, 631 
 
 
 
\bibitem[Kaspi et al.(2005)]{2005ApJ...629...61K} Kaspi, S., Maoz, D.,
  Netzer, H., Peterson, B.~M., Vestergaard, M., \& Jannuzi, B.~T.\
  2005, \apj, 629, 61 (K05)
 

 

\bibitem[Kellermann et al.(1989)]{1989AJ.....98.1195K} Kellermann, K.~I., 
Sramek, R., Schmidt, M., Shaffer, D.~B., \& Green, R.\ 1989, \aj, 98, 1195 

 

\bibitem[King(2008)]{2008NewAR..52..253K} King, A.\ 2008, \nar, 52, 253 

 

\bibitem[King et al.(2008)]{2008MNRAS.385.1621K} King, A.~R., Pringle, 
J.~E., \& Hofmann, J.~A.\ 2008, \mnras, 385, 1621 
 

\bibitem[Kurk et al.(2007)]{2007ApJ...669...32K} Kurk, J.~D., et al.\ 2007, 
\apj, 669, 32 

 
\bibitem[Kurosawa 
\& Proga(2009)]{2009MNRAS.397.1791K} Kurosawa, R., \& Proga, D.\ 2009, \mnras, 397, 1791 



\bibitem[Lawrence et al.(2007)]{2007MNRAS.379.1599L} Lawrence, A., et al.\ 
2007, \mnras, 379, 1599 


\bibitem[Lutz et al.(2010)]{2010ApJ...712.1287L} Lutz, D., et al.\ 2010, 
\apj, 712, 1287 

 
\bibitem[Marconi et al.(2004)]{2004MNRAS.351..169M} Marconi, A., Risaliti, 
G., Gilli, R., Hunt, L.~K., Maiolino, R., \& Salvati, M.\ 2004, \mnras, 351, 169 

 
\bibitem[Marconi et al.(2008)]{2008ApJ...678..693M} Marconi, A., Axon, 
D.~J., Maiolino, R., Nagao, T., Pastorini, G., Pietrini, P., Robinson, A., 
\& Torricelli, G.\ 2008, \apj, 678, 693 

 
 
\bibitem[Marziani et 
 al.(2009)]{2009A&A...495...83M} Marziani, P., Sulentic, J.~W., Stirpe, G.~M., Zamfir, S., \& Calvani, M.\ 2009, \aap, 495, 83 
 
 
\bibitem[McLure 
\& Dunlop(2004)]{2004MNRAS.352.1390M} McLure, R.~J., \& Dunlop, J.~S.\ 2004, \mnras, 352, 1390 

 
\bibitem[McLure 
\& Jarvis(2004)]{2004MNRAS.353L..45M} McLure, R.~J., \& Jarvis, M.~J.\ 2004, \mnras, 353, L45 
 
 
\bibitem[Merloni(2004)]{2004MNRAS.353.1035M} Merloni, A.\ 2004,
  \mnras, 353, 1035


\bibitem[Mineshige et al.(2000)]{2000PASJ...52..499M} Mineshige, S., 
Kawaguchi, T., Takeuchi, M., \& Hayashida, K.\ 2000, \pasj, 52, 499 

 
\bibitem[Miyaji et 
al.(2001)]{2001A&A...369...49M} Miyaji, T., Hasinger, G., \& Schmidt, M.\ 2001, \aap, 369, 49 
 
 
\bibitem[Netzer \& Trakhtenbrot(2007)]{2007ApJ...654..754N} Netzer, H., \& Trakhtenbrot, B.\ 2007, \apj, 654, 754 

 
\bibitem[Netzer et al.(2007)]{2007ApJ...666..806N} Netzer, H., et al.\ 
2007, \apj, 666, 806 (N07)
 
 

\bibitem[Netzer(2009a)]{2009ApJ...695..793N} Netzer, H.\ 2009, \apj, 695, 
793 

\bibitem[Netzer(2009b)]{2009MNRAS.399.1907N} Netzer, H.\ 2009, \mnras, 399, 
1907 


\bibitem[Netzer 
\& Marziani(2010)]{2010arXiv1006.3553N} Netzer, H., \& Marziani, P.\ 2010, arXiv:1006.3553 

 
\bibitem[Richards et al.(2006a)]{2006AJ....131.2766R} Richards, G.~T., et 
 al.\ 2006, \aj, 131, 2766 
 
 
\bibitem[Richards et al.(2006b)]{2006ApJS..166..470R} Richards, G.~T., et 
al.\ 2006, \apjs, 166, 470 
 
 
\bibitem[Salviander et al.(2007)]{2007ApJ...662..131S} Salviander, S., 
Shields, G.~A., Gebhardt, K., \& Bonning, E.~W.\ 2007, \apj, 662, 131 
 

\bibitem[Serjeant et 
al.(2010)]{2010A&A...518L...7S} Serjeant, S., et al.\ 2010, \aap, 518, L7 

 
\bibitem[Shankar(2009)]{2009NewAR..53...57S} Shankar, F.\ 2009, \nar, 53, 
57 


\bibitem[Shankar et al.(2004)]{2004MNRAS.354.1020S} Shankar, F., Salucci, 
P., Granato, G.~L., De Zotti, G., \& Danese, L.\ 2004, \mnras, 354, 1020 

 
\bibitem[Shankar et al.(2009)]{2009ApJ...690...20S} Shankar, F., Weinberg, 
D.~H., \& Miralda-Escud{\'e}, J.\ 2009, \apj, 690, 20 

%


\bibitem[Shankar et al.(2010a)]{2010ApJ...718..231S} Shankar, F., Crocce, 
M., Miralda-Escud{\'e}, J., Fosalba, P., 
\& Weinberg, D.~H.\ 2010, \apj, 718, 231 

\bibitem[Shankar et al.(2010b)]{2010MNRAS.401.1869S} Shankar, F., Sivakoff, 
G.~R., Vestergaard, M., \& Dai, X.\ 2010, \mnras, 401, 1869 


 


\bibitem[Shemmer et al.(2004)]{2004ApJ...614..547S} Shemmer, O., Netzer, 
H., Maiolino, R., Oliva, E., Croom, S., Corbett, E., 
\& di Fabrizio, L.\ 2004, \apj, 614, 547 (S04)

 

\bibitem[Shen et al.(2007)]{2007AJ....133.2222S} Shen, Y., et al.\ 2007, 
\aj, 133, 2222 




\bibitem[Shen et al.(2008)]{2008ApJ...680..169S} Shen, Y., Greene, J.~E., 
Strauss, M.~A., Richards, G.~T., \& Schneider, D.~P.\ 2008, \apj, 680, 169 

\bibitem[Shen et al.(2010)]{2010ApJ...719.1693S} Shen, Y., et al.\ 2010, 
\apj, 719, 1693 



 

\bibitem[Sigut 
\& Pradhan(2003)]{2003ApJS..145...15S} Sigut, T.~A.~A., \& Pradhan, A.~K.\ 2003, \apjs, 145, 15 

 
\bibitem[Sijacki et al.(2007)]{2007MNRAS.380..877S} Sijacki, D., Springel, 
V., Di Matteo, T., \& Hernquist, L.\ 2007, \mnras, 380, 877 

 
\bibitem[Silverman et al.(2008)]{2008ApJ...679..118S} Silverman, J.~D., et 
al.\ 2008, \apj, 679, 118 
 

\bibitem[Soltan(1982)]{1982MNRAS.200..115S} Soltan, A.\ 1982, \mnras, 200, 
115 


\bibitem[van der Bliek et al.(2004)]{2004SPIE.5492.1582V} van der Bliek, 
N.~S., et al.\ 2004, \procspie, 5492, 1582 
 

\bibitem[Vanden Berk et al.(2001)]{2001AJ....122..549V} Vanden Berk, D.~E., 
et al.\ 2001, \aj, 122, 549 

\bibitem[Vestergaard et al.(2008)]{2008ApJ...674L...1V} Vestergaard, M., 
Fan, X., Tremonti, C.~A., Osmer, P.~S., 
\& Richards, G.~T.\ 2008, \apjl, 674, L1 


\bibitem[Vestergaard 
\& Osmer(2009)]{2009ApJ...699..800V} Vestergaard, M., \& Osmer, P.~S.\ 2009, \apj, 699, 800 
 
 

\bibitem[Vestergaard \& Peterson(2006)]{2006ApJ...641..689V}
  Vestergaard, M., \& Peterson, B.~M.\ 2006, \apj, 641, 689


\bibitem[Vestergaard 
\& Wilkes(2001)]{2001ApJS..134....1V} Vestergaard, M., \& Wilkes, B.~J.\ 2001, \apjs, 134, 1 
 
 

 
\bibitem[Volonteri et al.(2003)]{2003ApJ...582..559V} Volonteri, M., 
Haardt, F., \& Madau, P.\ 2003, \apj, 582, 559 
 


\bibitem[Volonteri et al.(2005)]{2005ApJ...620...69V} Volonteri, M., Madau, 
P., Quataert, E., \& Rees, M.~J.\ 2005, \apj, 620, 69 

 
\bibitem[Volonteri(2010)]{2010A&ARv..18..279V} Volonteri, M.\ 2010, \aapr, 18, 279 


\bibitem[Wang 
\& Netzer(2003)]{2003A&A...398..927W} Wang, J.-M., \& Netzer, H.\ 2003, \aap, 398, 927 

 

\bibitem[Wang et al.(2009)]{2009ApJ...697L.141W} Wang, J.-M., et al.\ 2009, 
\apjl, 697, L141 

 
\bibitem[White et al.(1997)]{1997ApJ...475..479W} White, R.~L., Becker, 
R.~H., Helfand, D.~J., \& Gregg, M.~D.\ 1997, \apj, 475, 479 

\bibitem[White et al.(2008)]{2008MNRAS.390.1179W} White, M., Martini, P., 
\& Cohn, J.~D.\ 2008, \mnras, 390, 1179 

\bibitem[Wyithe 
\& Loeb(2009)]{2009MNRAS.395.1607W} Wyithe, J.~S.~B., \& Loeb, A.\ 2009, \mnras, 395, 1607 


 


\bibitem[Willott et al.(2010)]{2010AJ....140..546W} Willott, C.~J., et al.\ 
2010, \aj, 140, 546 


 
\bibitem[Woo 
\& Urry(2002)]{2002ApJ...581L...5W} Woo, J.-H., \& Urry, C.~M.\ 2002, \apjl, 581, L5 



 
\bibitem[York et al.(2000)]{2000AJ....120.1579Y} York, D.~G., et al.\ 2000, 
\aj, 120, 1579 
 



\end{thebibliography}
\end{document}